\documentclass[10pt]{iopart}

\expandafter\let\csname equation*\endcsname\relax
\expandafter\let\csname endequation*\endcsname\relax
\usepackage{amsmath} 
\usepackage{amssymb}
\usepackage[utf8]{inputenc}
\usepackage[english]{babel}
\usepackage{hyperref}
\usepackage{graphicx}
\usepackage{subcaption}
\usepackage{color}
\usepackage[dvipsnames]{xcolor}
\usepackage[normalem]{ulem}
\usepackage{bm} 

\newcommand{\II}{I\!I}

\begin{document}
\title{Theoretical framework for pairwise microswimmer interactions}
\author{Sebastian Ziegler\textit{$^{a}$}, Thomas Scheel\textit{$^{ab}$}, Maxime Hubert\textit{$^a$}, Jens Harting\textit{$^{bc}$} and Ana-Sunčana Smith$^{\ast}$\textit{$^{ad}$}}
\address{$^{a}$~PULS Group, Department of Physics and Interdisciplinary Center for Nanostructured Films, Friedrich-Alexander-University Erlangen-Nuremberg, Cauerstr. 3, 91058 Erlangen, Germany}
\address{$^{b}$~Helmholtz Institute Erlangen-Nuremberg for Renewable Energy (IEK-11), Forschungszentrum Jülich, Fürther Str. 248, 90429 Nuremberg, Germany}
\address{$^{c}$~Department of Chemical and Biological Engineering and Department of Physics, Friedrich-Alexander-University Erlangen-Nuremberg, Fürther Str. 248, 90429 Nuremberg, Germany}
\address{$^{d}$~Group for Computational Life Sciences, Division of Physical Chemistry, Ruđer Bošković Institute, Bijeni\v{c}ka cesta 54, 10000 Zagreb, Croatia}
\ead{$^\ast$smith@physik.fau.de}

\begin{abstract}
Hydrodynamic interactions are crucial for determining the cooperative behavior of microswimmers at low Reynolds numbers. Here we provide a comprehensive analysis of the scaling and strength of the interactions in the case of a pair of three-sphere swimmers with intrinsic elasticity. Both stroke-based and force-based microswimmers are analyzed using an analytic perturbative approach. Following a detailed analysis of the passive interactions, as well as active translations and rotations, we find that the mapping between the stroke-based and force-based swimmers is only possible in a low driving frequency regime where the characteristic time scale is smaller than the viscous one. Furthermore, we find that for swimmers separated by up to hundreds of swimmer lengths, swimming in pairs speeds up the self propulsion, due to the dominant quadrupolar hydrodynamic interactions. Finally, we find that the long term behavior of the swimmers, while sensitive to initial relative positioning, does not depend on the pusher or puller nature of the swimmer.

\end{abstract}

\maketitle

\section{Introduction}
Locomotion of microscopic organisms such as bacteria or sperms is governed by laws which are different from those governing the world tangible by humans. Due to the dominance of viscous drag over inertia and the ensuing time-independence of the Stokes equations, a successful swimming strategy has to break the time-reversal symmetry \cite{Purcell1977}. Several theoretical models \cite{Blake1971, Felderhof2006, EarlPooleyRyder2007, NajafiGolestanian2004, AvronKennethOaknin2005, DowntonStark2009,  PandeSmith2015, ZieglerHubertVandewalle2019, Wang2019, DaddiLisickiHoell2018, DaddiKurzthalerHoell2019, RizviFarutinMisbah2018}, experimental realizations \cite{ DreyfusBaudryRoper2005, AhmedLuNourhani2015, GrosjeanLagubeauDarras2015, GrosjeanHubertLagubeau2016, GrosjeanHubertCollard2018, ZhengDaiWang2017, HamiltonPetrovWinlove2017, BryanShelleyParish2017, HamiltonGilbertPetrov2018, Collard2020} and simulations \cite{BeckerKoehlerStone2003, ZottlStark2012, PicklGoetzIglberger2012, SukhovZieglerXie2019, PicklPandeKoestler2017} have been employed to scrutinize the details of locomotion under these laws. Overall, due to the absence of inertia, the force to self-propel exerted by a microswimmer on its surrounding  is balanced by friction, and force monopoles are not present. Consequently, a self-propelled swimmer typically induces a dipolar flow field, allowing to classify swimmers as pushers or pullers. In addition to it, higher order flow fields are induced \cite{SpagnolieLauga2012}, which, nonetheless, induce hydrodynamic interactions between distant swimmers. Interestingly, living microswimmers are capable of not only sensing these flows, but also reacting to them, which is associated with important physiological functions \cite{Shen2012, Guasto2011, Uppaluri2012, Mathijssen2019}.

Hydrodynamic interactions have been investigated for a number of model microswimmers \cite{PooleyAlexanderYeomans2007, AlexanderPooleyYeomans2008, AlexanderPooleyYeomans2009, FarzinRonasiNajafi2012, KurodaYasudaKomura2019, IshikawaSimmondsPedley2006, MirzakhanlooJalaliAlam2018}, leading to a consensus that, in a first approximation, a swimmer can be considered as a passive entity subject to the average flow produced by nearby swimmers. With this approximation as a starting point, suspensions of many swimmers have been investigated using simulations \cite{Bardfalvy2020, EvansIshikawaYamaguchi2011} and theoretical approaches \cite{Hoell2018, Reinken2018}.
Additionally, at the single swimmer level an active component to the interactions, depending on the time-resolved stroke of both swimmers, has been reported throughout the literature \cite{PooleyAlexanderYeomans2007, AlexanderPooleyYeomans2009, FarzinRonasiNajafi2012, KurodaYasudaKomura2019}. In opposition to the expectation that passive interactions are sufficient, it has been reported that the active component actually dominates for swimmer separations smaller than a threshold value depending on the details of the stroke \cite{PooleyAlexanderYeomans2007}.
However, several competing models for the active interactions are currently discussed, providing different predictions for the scaling and the sign of the interaction effects \cite{PooleyAlexanderYeomans2007, AlexanderPooleyYeomans2009, FarzinRonasiNajafi2012}. Given that most models rely on the same basic assumptions, the origins of these differences have not been clearly established thus far.  

Most of these models consider interacting bead-based microswimmers subject to a prescribed swimming stroke. Yet, there is a second family of bead-based microswimmers where the swimmer arms are replaced with elastic springs, and oscillating external driving forces induce the swimming stroke \cite{Felderhof2006, PandeSmith2015, ZieglerHubertVandewalle2019}. These so-called bead-spring swimmers can adapt their swimming stroke and velocity to the surrounding fluid \cite{PandeMerchantKrueger2017a} and also to the presence of other swimmers, an effect not taken into account in previous works. For a single swimmer a mapping from prescribed forces to the corresponding stroke exists, which reproduces the same swimming velocity and flow field in both approaches. Consequently, a natural question to ask is if the same correspondence applies for the interaction of two swimmers. 

In this paper, we answer this question by calculating the interaction of two linear three-bead microswimmers in both the force-based (FB) and stroke-based (SB) models using a recently developed perturbative approach \cite{ZieglerHubertVandewalle2019}. 
While passive interactions are equivalent in the stroke-based and force-based models, the additional active component to the interactions, depending on the swimming stroke of both species, is in general not equivalent in both approaches. Interestingly, the interactions between force-based swimmers becomes equivalent to the interactions between stroke-based swimmers if the driving frequency in the force-based model is small as compared to the inverse viscous time of the system. Hence, the interactions of stroke-based swimmers should actually be considered as special case of the interactions of two force-based swimmers. These findings are systematically discussed in the context of the existing literature \cite{PooleyAlexanderYeomans2007,AlexanderPooleyYeomans2009, FarzinRonasiNajafi2012, KurodaYasudaKomura2019}. 

The analysis of the scaling laws is complemented by an in-depth discussion of the cooperative effects between swimmers. Unlike previously observed, we show that two collinear or side-by-side swimmers with parallel swimming direction typically benefit mutually from each other due to the time-averaged flow fields they produce. In simple words, a swimmer propagates faster in a pair compared to when it is alone, even if it is leading. This result, which is a consequence of vanishing Reynolds numbers, is independent of the details of the swimming stroke due to a close relation of the leading order swimming velocity and flow fields produced by each swimmer.

The remainder of the article is structured as follows: We first introduce both the force-based and stroke-based swimmer models, elaborate on the perturbative approaches and establish the mapping between both models. We then proceed in section \ref{sec:results} to calculate the average flow fields of our swimmers. Focusing on a pair of side-by-side or trailing swimmers, we then investigate the passive interactions and, consequently, the active translations and rotations, comparing the obtained results to findings in the literature. Finally, we expand our analysis to arbitrary positioning of the swimmers, and estimate the long-term behavior of a pair. The most important findings are summarized in section \ref{sec:conclusion}, which concludes the paper. Details on the perturbative calculations for the force-based and stroke-based models as well as for the proportionality of the leading order swimming velocity and flow fields are given in the Appendices A-D.

\section{Model}
\label{sec:Model}

\begin{figure}
    \begin{center}
        \includegraphics[scale=0.75]{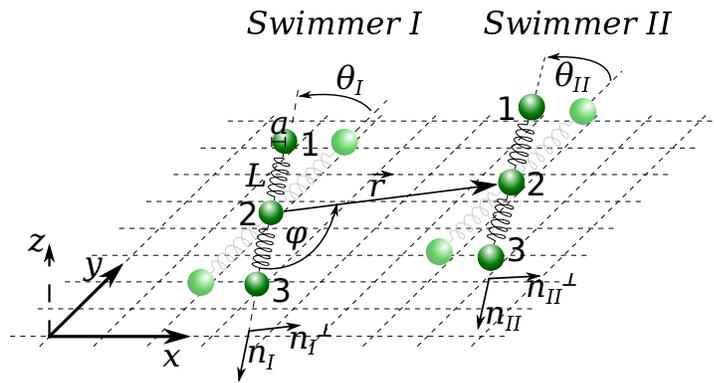}
        \caption{Sketch of two interacting in-plane linear three-sphere swimmers.}
        \label{fig_1}
    \end{center}
\end{figure}

\subsection{Force-based model of the three-bead swimmer}
\label{sec:forcebasedmodel}

We consider two linear three-bead swimmers identified by subscripts $s, p \in \{I, \II \}$. We restrict our analysis to a two-dimensional configuration space ($d = 2$), in our case the $x$-$y$-plane, as depicted in figure \ref{fig_1}. Each swimmer consists of three identical spherical beads of radius $a$, denoted by subscripts $i, j, m \in \{1,2,3\}$. The beads are connected by identical linear harmonic springs of spring constant $k$ and equilibrium length $L$, such that the spring force between two connected beads $si$ and $sj$, acting on the latter, is given by 
\begin{equation}
    \vec{g}(\vec{x}) := - k (|\vec{x}| - L) \cdot \frac{\vec{x}}{|\vec{x}|},
\end{equation} 
with $\vec{x} = \vec{R}_{sj} - \vec{R}_{si}$ the vector connecting both beads. We employ a convenient double index notation to identify the beads, where the first index corresponds to the swimmer and the second index to the bead number within this swimmer. 

In this work we only consider swimmers with a rigid joint connecting both swimmer arms, i.e. the two swimmer arms cannot bend with respect to each other. This allows to systematically compare force-based to linear stroke-based swimmers \cite{NajafiGolestanian2004, PooleyAlexanderYeomans2007, AlexanderPooleyYeomans2009, FarzinRonasiNajafi2012}, to which this constraint is inherent. 

As shown in figure \ref{fig_1}, swimmer $I$ is oriented along $\vec{n}_I = (\sin \theta_I, -\cos \theta_I)$, swimmer $\II$ along $\vec{n}_{\II} = (\sin \theta_{\II}, -\cos \theta_{\II})$ and the middle beads of both swimmers are separated by $\vec{r} = r (\sin (\varphi + \theta_I), -\cos (\varphi + \theta_I))$. Here, $\varphi$ denotes the angle enclosed between $\vec{n}_I$ and $\vec{r}$. The swimmers are immersed in a viscous fluid of viscosity $\eta$, where the Reynolds number of a single bead is assumed to be zero. The interaction between two spherical beads at sufficiently large distance is then described by the Oseen tensor 
\begin{equation}
    \hat{T}(\vec{x}) := \frac{1}{8 \pi \eta |\vec{x}|} \left( \hat{1} + \frac{\vec{x} \otimes \vec{x} }{|\vec{x}|^2} \right),
\end{equation}
with $\hat{1}$ the unit matrix and $\otimes$ the tensor product. 

Each swimmer $s$ is driven by sinusoidal forces of frequency $\omega$ acting on the two outer beads (bead s1 and s3) along the respective adjacent arm with relative phase shift $\alpha_s$. The driving forces on swimmer $\II$ precede those on swimmer $I$ by a phase shift of $\gamma$: 
\begin{eqnarray}
    \vec{E}_{I1} (t) = A_{I1} \sin(\omega t) \vec{n}_I, \ \vec{E}_{I3} (t) = A_{I3} \sin(\omega t + \alpha_I) \vec{n}_I, \nonumber \\   \vec{E}_{\II1} (t) = A_{\II1} \sin(\omega t + \gamma) \vec{n}_{\II}, \ \vec{E}_{\II3} (t) = A_{\II3} \sin(\omega t + \gamma + \alpha_{\II}) \vec{n}_{\II},
    \label{eq:drivingForces}
\end{eqnarray} 
with $\vec{E}_{si}$ the force on the bead $si$, $A_{si}$ the respective force amplitude and $t$ the time. 
We introduce a second set of independent variables for the driving forces which is required in the perturbative calculation, $A := A_{I1}, B_I := A_{I2}/A_{I1}, C := A_{\II 1}/A_{I1}, B_{\II} := A_{\II 2}/ A_{\II 1}$, with $A$ the overall strength of the driving forces and $B_I, B_{\II}, C$ dimensionless relative parameters.
The force on each middle bead $s2$ is determined by requesting the sum of all three forces on the beads of swimmer $s$ to vanish \cite{PandeSmith2015}, $\vec{E}_{s2}(t) := -\vec{E}_{s1} (t) - \vec{E}_{s3} (t)$, as a prerequisite for self-propulsion.

We then devise an equation of motion for all of the $n = 6$ beads,
\begin{eqnarray}
\fl
    \frac{d}{dt} \vec{R}_{si} (t) = & \mu \left( \vec{E}_{si} (t) + \sum_{j \in \mathrm{NN}(i)} \vec{g}(\vec{R}_{sj}(t) - \vec{R}_{si}(t)) \right) + \nonumber \\
    \fl
    & \sum_{(pj) \neq (si)} \hat{T}(\vec{R}_{si}(t) - \vec{R}_{pj}(t)) \cdot \left( \vec{E}_{pj} (t) + \sum_{m \in \mathrm{NN}(j)} \vec{g}(\vec{R}_{pm}(t) - \vec{R}_{pj}(t)) \right),
    \label{eq:EOM}
\end{eqnarray}
with $\vec{R}_{si}$ the position of the respective bead and $\mu = (6 \pi \eta a)^{-1}$ the Stokes mobility. $\mathrm{NN}(j)$ denotes all beads connected to bead $j$ via a spring within the geometry of the linear swimmer, and in the remaining summation at the beginning of the second line we sum over all beads $pj$ different from the bead $si$. 
To obtain the swimmer behavior analytically, we employ a recently developed perturbative approach to bead-spring microswimmers \cite{ZieglerHubertVandewalle2019} (details in \ref{ch:appendixB}). Also, we solve \eqref{eq:EOM} numerically using the NDSolve function from Mathematica \cite{Mathematica2017}, with a superimposed angle spring potential ensuring the rigidity of the joint at the middle bead of each swimmer (details in \ref{ch:appendixA}). 

A system of bead-spring swimmers as described above can be rescaled using the bead radius $a$ as the length unit and the viscous time $t_V := 6 \pi \eta a/k$ as the time unit, in order to identify the effective parameters \cite{ZieglerHubertVandewalle2019}. We define the dimensionless parameters $q:= a/r$ and $\nu := a/L$ encoding together with $\varphi, \theta_1$ and $\theta_2$ the swimmer geometry, and the rescaled driving frequency $\Gamma := \omega t_V$, comparing the two time scales set by the driving forces and the viscous time. 

In the subsequent analysis, we will illustrate the results for the swimmer interactions using a swimmer of puller- and a swimmer of pusher-type with distinct parameters. For the puller, we choose 
\begin{equation}
    \frac{a}{L} = \frac{1}{6}, \ A_{s1} = A_{s3} = \frac{5}{4} ka, \ \beta_s = \frac{\pi}{2}, \ \omega t_V = 0.7402
\end{equation}
and for the pusher-type swimmer
\begin{equation}
    \frac{a}{L} = \frac{1}{6}, \ A_{s1} = \frac{5}{3} k a, \ A_{s3} = \frac{5}{6} ka, \ \beta_s = \frac{\pi}{2}, \ \omega t_V = 0.7402.
\end{equation}

\subsection{Stroke-based model of the three-bead swimmer}
\label{ch:strokeBasedModel}
In the stroke-based model, we consider two swimmers with identical geometry as in the force-based model. In contrast to the force-based swimmer, the length of each swimmer arm is directly prescribed as a function of time
\begin{eqnarray}
    L_{I1} (t) &= L + \xi_{I1} \sin( \omega t), \ L_{I2} (t) = L + \xi_{I2} \sin( \omega t + \beta_I), \nonumber \\ L_{\II1}(t) &= L + \xi_{\II1} \sin(\omega t + \delta), \ L_{\II2} = L + \xi_{\II2} \sin(\omega t + \delta + \beta_{\II}), 
\end{eqnarray}
with $L$ the average arm length, $\xi_{sb}$ the corresponding arm oscillation amplitude (indices for the swimmer arms $b,c \in \{1, 2\}$) and $\beta_I$, $\beta_{\II}$ and $\delta$ the phase shifts within each swimmer and between both, respectively. 
Similarly to the force-based model, we introduce a second set of parameters for the arm oscillation amplitudes required in the perturbative analysis, $\xi := \xi_{I1}, D_I := \xi_{I2}/\xi_{I1}, F := \xi_{\II 1}/\xi_{I1}, D_{\II} := \xi_{\II 2}/\xi_{\II 1}$, with $\xi$ the overall amplitude of the arm oscillations.
With both swimmers constrained to linear shape and to the prescribed arm lengths, the system contains six undetermined degrees of freedom, which we choose to be the positions of both middle beads, $\vec{R}_{I2}$ and $\vec{R}_{\II 2}$, as well as the orientations of both swimmers, $\theta_I$ and $\theta_{\II}$. The positions of all other beads are then given by the prescribed arm lengths and the constraint on the linear swimmer shape. 

The relation of the bead velocities to the forces on the beads is given similarly as in the force-based model by
\begin{equation}
    \frac{d}{dt} \vec{R}_{si} (t) = \mu \, \vec{F}_{si} (t) + \sum_{(pj) \neq (si)} \hat{T} \left(\vec{R}_{si}(t) - \vec{R}_{pj}(t)\right) \cdot \vec{F}_{pj}(t),
    \label{eq:StokesLaw}
\end{equation}
where $\vec{F}_{si} (t)$ denotes the forces which are acting on the beads to enforce the prescribed strokes. 
The remaining six degrees of freedom are then determined by the condition that the total vectorial force on each swimmer vanishes, 
\begin{equation}
\sum_{i = 1}^3 \vec{F}_{si}(t) = 0, \ s \in \{I, \II\},
\label{eq:forceFreeCond}
\end{equation}
 as well as the total torque on each swimmer, which is a scalar quantity in the 2D framework employed here:
\begin{equation}
(- L_{s1} (t) \vec{F}_{s1}(t) + L_{s2} (t) \vec{F}_{s3}(t)) \cdot \vec{n}^\perp_s = 0, \ s \in \{I, \II\}.
\label{eq:torqueFreeCond}
\end{equation}

Using the compact notation $\bm{R} = (\vec{R}_{I1}, \vec{R}_{I2}, \vec{R}_{I3}, \vec{R}_{\II 1}, \vec{R}_{\II 2}, \vec{R}_{\II 3})$ and $\bm{F} = (\vec{F}_{I1}, ..., \vec{F}_{\II 3})$, we can re-express \eqref{eq:StokesLaw} as 
\begin{equation}
\frac{d}{dt} \bm{R} = \underline{\mu}(\bm{R}) \, \bm{F}. 
\label{eq:StokesLawCompact}
\end{equation}
Here, $\underline{\mu} (\bm{R})$ denotes the $(n \cdot d) \times (n \cdot d)$-dimensional mobility matrix \cite{Dhont1996, ZieglerHubertVandewalle2019} and we neglect the time-dependence of $\bm{R}$ and $\bm{F}$ in our notation for the sake of brevity. Vectors in the $(n \cdot d)$-dimensional configuration space of all bead positions are denoted by bold symbols and higher order tensors on this space by underlined symbols.
In order to perturbatively calculate the swimmer behavior, we expand the undetermined variables $\vec{R}_{I2}, \vec{R}_{\II 2}, \theta_I$ and $\theta_{\II}$ as power series in $\xi$ and $q$, and from this calculate an expansion of $d \bm{R} /dt$ and $\underline{\mu}(\bm{R})$. By also expanding \eqref{eq:StokesLawCompact}, we are able to express the components of $\bm{F}$ associated to different powers of $\xi$ and $q$ in terms of the middle bead positions and swimmer orientations. The force-free and torque-free conditions, \eqref{eq:forceFreeCond} and \eqref{eq:torqueFreeCond}, close the equations obtained and by solving them we obtain the full swimmer behavior as an expansion in $\xi$ and $q$. The perturbative calculation is explained in more detail in \ref{ch:appendixC2}.

\subsection{Mapping from force-based to stroke-based model}
For a single force-based swimmer, the swimming velocity can be either calculated directly in the force-based framework \cite{ZieglerHubertVandewalle2019}, or one can equivalently extract the time-dependent arm lengths and insert them in the respective expression for the swimming velocity in the stroke-based model \cite{GolestanianAjdari2008}. Since the single swimmer behavior is uniquely defined by the time-dependent arm lengths together with the force-free condition \cite{GolestanianAjdari2008}, both ways must yield the same result. This means that it exists a mapping from the parameter space of the force-based system to the parameter space of the stroke-based system which preserves the overall swimmer dynamics. For the sake of simplicity, we restrict the mapping to the first order in $A$ and $\xi$ in the subsequent calculations, where the average arm lengths in the force-based model are still equal to the spring lengths in the mechanical equilibrium, $L$ \cite{ZieglerHubertVandewalle2019}. Hence, we assume the geometry of the force-based swimmer in its mechanical equilibrium for the time-averaged shape of the stroke-based swimmer, i.e. we assume average arm lengths $L$, and only require to map the swimming stroke, i.e. the arm oscillation amplitudes. This reproduces consistent second order swimming velocities and flow fields in both models and is hence sufficient for the subsequent comparison.
\begin{figure*}
    \centering
	\includegraphics[scale=0.85]{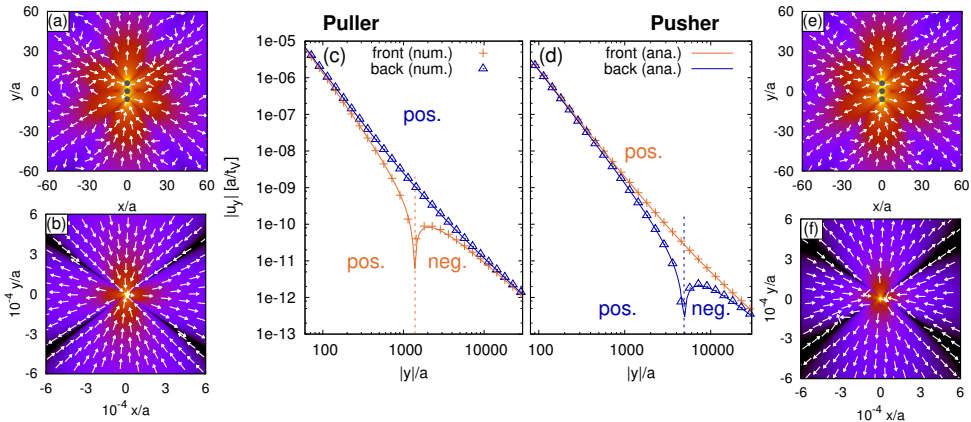}
    \caption{Flow fields of a puller- and pusher-type swimmer oriented along the $y$-axis. Intermediate field, far field and flow field amplitude on the y-axis for the puller-like swimmer ((a), (b) and (c), respectively) and for the pusher-like swimmer ((e), (f) and (d), respectively). In both cases, the swimmer propagates in positive $y$-direction. In (c) and (d), the flow amplitude in the front of the swimmer ($y > 0$) is given by orange pluses and in the back of the swimmer ($y < 0$) by blue triangles, both obtained numerically. Solid lines in the respective colors denote the corresponding analytical results. The swimmer parameters are chosen as defined in section \ref{sec:forcebasedmodel}.}
    \label{fig_2}
\end{figure*}

The inverse map from the stroke-based to the force-based system is not unique, since the force-based model comprises one additional parameter, namely the spring constant $k$. It gives rise to the viscous time scale $t_V$ and the rescaled driving frequency $\Gamma$, with respect to which a force-based swimmer attains the maximum driving speed at $\Gamma \approx 1$ \cite{SukhovZieglerXie2019}, and is unable to self-propel for $\Gamma \gg 1$ and $\Gamma \ll 1$ \cite{PandeMerchantKrueger2017a}. In contrast, $\omega$ has no impact on the propulsion of a stroke-based swimmer measured over one stroke \cite{GolestanianAjdari2008} and drops out in a suitable rescaling. This is because with a prescribed stroke, calculating the swimmer behavior reduces to a purely geometric problem due to the time-independence of the Stokes equations \cite{GolestanianAjdari2008}.

The natural question arising then is: Is also the interaction of two or more swimmers equivalent in the force- and stroke-based models under the above mapping, or does the interaction in the force-based model change with $\Gamma$, when the driving forces on each swimmer are such that the stroke amplitude and phase of each swimmer alone would be constant?

\section{Results and discussion}
\label{sec:results}

\subsection{Flow field of a single swimmer}
A single linear three-sphere swimmer with equal bead radii typically produces a flow field which is dominantly dipolar in the very far field and quadrupolar at intermediate distances (figure \ref{fig_2}) \cite{PooleyAlexanderYeomans2007}. The transition between both fields depends on the swimmer parameters, and appears in our case at the order of $100 \, L$. 
The time-averaged dipolar flow field scales to leading order as $A^4$ or $\xi^4$, depending on the model considered, and is given in terms of the swimming stroke as
\begin{eqnarray}
\fl
    \vec{u}^\mathrm{dip}(\vec{r}) = \frac{\omega  a^2}{2 r^2} \frac{\xi_1 \xi_2 \left(16947 a^3-24924 a^2 L+11664 a L^2-1856 L^3\right) \left(\xi_1^2-\xi_2^2\right)}{ (3 a-4 L)^3 (7 a-4 L)^3} \times \nonumber \\ \sin (\beta) \frac{1}{2} \left[3 \frac{(\vec{r} \cdot \vec{n})^2}{r^2} - 1 \right] \frac{\vec{r}}{r},
    \label{eq:dipolarFlowField}
\end{eqnarray}
with $r := |\vec{r}|$ and $\vec{n}$ the unit vector along the swimmer axis (details on the calculation in \ref{ch:appendixC1}). We have omitted the swimmer index in $\xi_1$ and $\xi_2$ since this result holds for a single swimmer. 
In contrast, the average quadrupolar flow field scales to leading order as $A^2$ or $\xi^2$ and reads in terms of the stroke
\begin{eqnarray}
\fl
    \vec{u}^\mathrm{quad}(\vec{r}) = -\frac{\omega  a^2}{r^3}\frac{\xi_1 \xi_2 L^2 (147 a-68 L) \sin (\beta )}{(3 a-4 L) (7 a-4 L)^2} \times \nonumber \\ \frac{1}{4} \left[ 3 \frac{(\vec{r} \cdot \vec{n})}{r} \left( 5 \frac{(\vec{r} \cdot \vec{n})^2}{r^2} - 3 \right) \frac{\vec{r}}{r} - \left( 3 \frac{(\vec{r} \cdot \vec{n})^2}{r^2} - 1 \right) \vec{n} \right]. 
    \label{eq:quadrupolarFlowField}
\end{eqnarray}
Using the mapping from the parameters of the force-based to the stroke-based model at sufficient order in $\xi$ and $A$, one obtains similar expressions for the force-based swimmer, which we do not include as they would become too large. 

To understand the predominance of the quadrupolar regime in the average flow field up to distances of hundreds of swimmer lengths, we first consider the special case of a swimmer with $\xi_{1} = \xi_{2}$, i.e. equal arm oscillation amplitudes. In this case, the swimmer becomes invariant under a combined time-reversal ($t \to -t$) and parity ($\vec{r} \to -\vec{r}$) transformation (TP transformation) and the dipolar regime in the time-averaged flow field is lost \cite{PooleyAlexanderYeomans2007}. Indeed, flow fields which decay with an even order in the distance $r$ from the swimmer (dipolar, octopolar, ...) are not consistent with TP invariance. Thus, the far field of a TP invariant and self-propelled swimmer is quadrupolar, decaying as $r^{-3}$, and the dipolar regime in figure \ref{fig_2} would disappear. However, even a TP-invariant swimmer still produces a non-zero dipolar flow field that oscillates with the frequency of the swimming stroke. 

For $\xi_{1} \neq \xi_{2}$, which is the case for both the puller- and pusher-type swimmer defined in section \ref{sec:Model}, the TP invariance is broken and a non-zero time-averaged dipolar flow field arises. Notably, the leading order, $\xi^2$ contribution to all even-order time-averaged flow fields still vanishes, and the first non-zero contribution arises at order $\xi^4$. The reasoning to show this extends an argument made originally by Golestanian and Ajdari \cite{GolestanianAjdari2008}. Similarly to the time-averaged swimming velocity, the amplitude of each component (dipolar, quadrupolar, ...) of the average $\xi^2$ flow field must be given by a geometric prefactor times the area in the configuration space enclosed by the swimmer's stroke (\ref{ch:appendixD}). First, this leads to the important conclusion that, at order $\xi^2$, average flow field and swimming velocity are proportional and linked by a factor depending on the average swimmer geometry only. Accordingly, knowing the swimmer's geometry, one can, at leading order, directly infer the average flow field from the swimming velocity. Second, we conclude that the proportionality coefficient needs to vanish for all even-order components (dipolar, octopolar, ...) of the average flow field, since it vanishes for the special case $\xi_{1} = \xi_{2}$ and is generic for linear three-bead swimmers with equal radii and equal average arm lengths. 

In the case of sinusoidal driving forces with a single frequency, also the $\xi^3$ time-averaged flow field vanishes since it is associated to odd products of the sinusoidal stroke, which averages to zero over a stroke cycle. This explains why passive dipolar interactions therefore arise firstly at fourth order in the stroke amplitude ($\sim \xi^4/r^2$) \eqref{eq:dipolarFlowField}. 
In \cite{PooleyAlexanderYeomans2007}, a stroke comprising stepwise arm contractions at constant velocity \cite{NajafiGolestanian2004} has been employed, associated to infinitely many harmonics of the base frequency $\omega$ in a Fourier decomposition. Therefore, a non-zero average dipolar flow field has already been observed at third order in $\xi$, making the quadrupolar regime less dominant than in the case of sinusoidal driving with a single pure frequency. 
\begin{figure*}[h]
    \centering
	\includegraphics[scale=0.9]{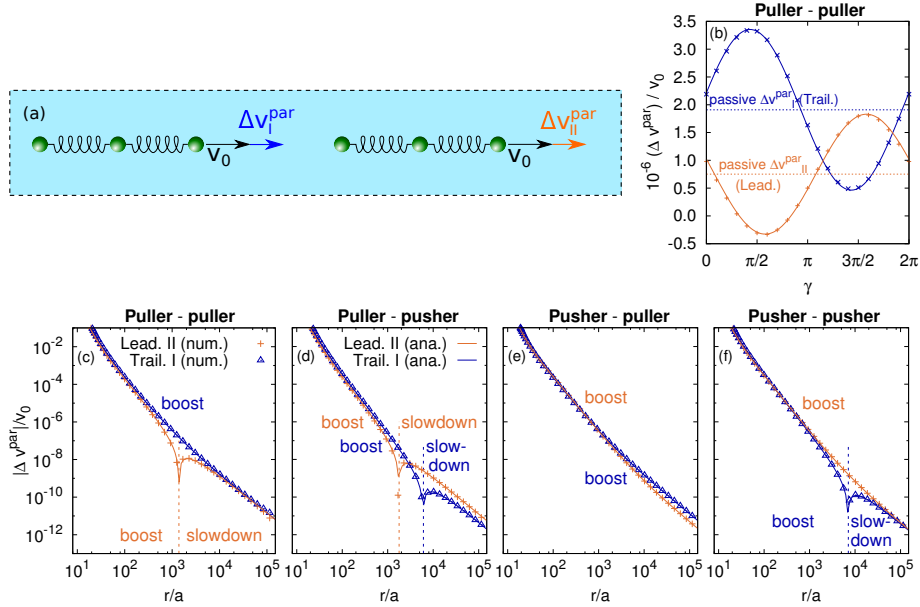}
    \caption{Behavior of two linear three-bead swimmers interacting collinearly. (a) Sketch. In (b), the dependence of the boost on the phase shift $\gamma$ between both swimmers at distance $r = 500a$ is shown for two pullers. Panels (c) - (f) show the increase/decrease in the swimming velocity for all four combinations of pullers and pushers with the phase shift between both swimmers fixed to $\gamma = 0$. The headlines of each plot indicate the swimmer types with first the trailing and second the leading swimmer. Numerical results are shown by orange pluses for the leading and by blue triangles for the trailing swimmer. The solid lines in respective color represent the analytical results.}
    \label{fig_3}
\end{figure*}

\subsection{Passive interaction}
While a single swimmer alone would propagate with velocity $v_0$ along its axis, in the presence of another swimmer it experiences both additional translation and rotation due to the hydrodynamic interactions. We split the swimmer's velocity into a component $v_0 + \Delta v_\mathrm{par}$ parallel to its axis $\vec{n}_s$ and into a component $v_\mathrm{ort}$ along $\vec{n}^\perp_s$ (see figure \ref{fig_1}). Additionally, the swimmer rotates with an angular velocity $\Omega$ which is defined as positive when the swimmer rotates counterclockwise. We restrict our subsequent analysis to the velocities time-averaged over one swimming cycle. 

We distinguish between active and passive interaction effects by their dependence on the phase shift between both swimmers, $\gamma$ in the force-based model and $\delta$ in the stroke-based model. Considering the velocity of say swimmer $I$, passive terms are constant with respect to $\gamma$ and $\delta$. In contrast, active interactions explicitly depend on the phase shift between both swimmers. 

In agreement with previous work we find that all passive interactions, in particular both translation and rotation, are solely due to the average flow field that swimmer $\II$ would produce if it was alone in an otherwise empty and unperturbed fluid \cite{PooleyAlexanderYeomans2007, AlexanderPooleyYeomans2009}. This result holds for both the force-based and the stroke-based model. Note that for the passive interactions the single-swimmer flow field of swimmer $\II$ is relevant, and not the average flow field produced by swimmer $\II$ in the presence of swimmer $I$. Since the difference between the two situations arises from the swimming activity of swimmer $I$, this effect gives rise to active interactions. Thus, passive interactions are, via their definition, equivalent in the force-based and stroke-based models, i.e. they agree if the force protocol and stroke are chosen such that they produce the same single swimmer behavior.  
The passive translation of swimmer $I$ equals the local average flow field $\vec{u}(\vec{x})$ produced by swimmer $\II$. The calculation of passive rotation is more involved,
\begin{equation}
    \Omega^\mathrm{pass.}_I = \dot{\theta}^\mathrm{pass.}_{I} =  (\vec{n}_s \cdot \nabla) \left[\vec{n}^\perp_s \cdot \vec{u}(\vec{x}) \right]|_{\vec{x} = \vec{R}_{I 2}},
\end{equation}
as it results from the interplay of the swimmer shape and the local gradient of the average flow field. 

\begin{figure*}
\centering
	\includegraphics[scale=0.9]{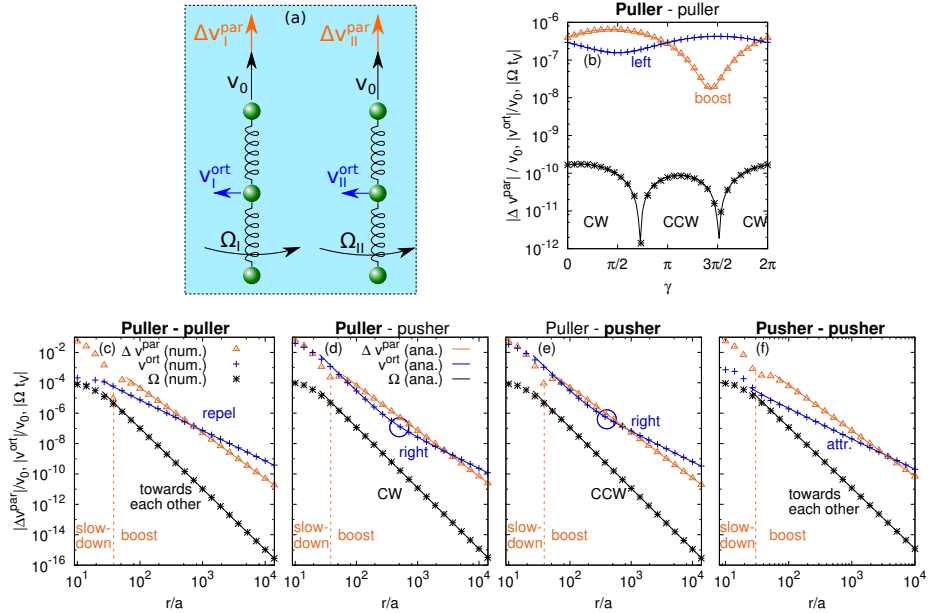}
\caption{Behavior of two linear three-bead swimmers in the side-by-side configuration. (a) Sketch. In (b), the dependence of all three velocities on the phase shift $\gamma$ between both swimmers is shown for two pullers at distance $r = 500a$. 
Panels (c) - (f) show the boost/slowdown along the swimmer axis which scales as $r^{-3}$ at large $r$ ($\mathrm{v^{par}}$, orange triangles), the induced velocity orthogonal to it which scales as $r^{-2}$ at large $r$ and shows a transition to $r^{-4}$ scaling in (d), (e) marked by blue circles ($\mathrm{v^{ort}}$, blue pluses) and the angular velocity $\Omega$ which scales $r^{-4}$ at large $r$ (black stars) for all four combinations of pullers and pushers. The phase shift between both swimmers is fixed to $\gamma = 0$. For equal swimmer types and $\gamma = 0$, the swimmers behave symmetrically and the values plotted hold for both swimmers (attr.: attractive), for different swimmer types the values are plotted for the swimmer highlighted in bold in the headline (CW: clockwise, CCW: counterclockwise). }
\label{fig_4}
\end{figure*}

Mediated by the passive interactions, the transition from predominantly quadrupolar to dipolar with increasing distance is also found in the interaction of two collinear swimmers, i.e. swimmers with a common axis ($\theta_1 = \theta_2 = 0, \varphi = 0$) (figure \ref{fig_3}). For side-by-side swimmers with parallel swimming direction, defined by $\theta_1 = \theta_2 = 0$ and $\varphi = \pi/2$, passive dipolar interaction effects are found only in $v^\mathrm{ort}$ and passive quadrupolar interactions only in $v^\mathrm{par}$ (figure \ref{fig_4}). This is due to the shapes of the respective flow fields (figure \ref{fig_2}). In the general case of two interacting swimmers, also passive quadrupolar rotation would be observed, resulting from the gradient of the dipolar time-averaged flow field. However, in the two most simple swimmer configurations chosen here for illustration, this term is absent due to the symmetries of the dipolar flow field. 

The proportionality of the $\xi^2$ or $A^2$ time-averaged flow field and the swimming velocity has important implications for the swimmer interaction. Namely, for swimmers of given geometry, the passive interactions between the swimmers depend only on their swimming velocities and their relative positioning, but not on the details of the strokes. The factor of proportionality for the quadrupolar flow field of a single swimmer, measured in front of the swimmer along the swimmer axis $\vec{n}_s$, relative to the swimming velocity is given by
\begin{equation}
(\vec{u}^\mathrm{quad} (r \vec{n}_s) \cdot \vec{n}_s)/v_0 = \frac{3 a L^3 (68 L - 147 a)}{r^3 \left(56 L^2 -198 a L + 189 a^2\right)} \geq 0.
\end{equation}
This ratio is positive for values $L \geq 6a$ where the Oseen approximation can be expected to be sufficiently good. 
Therefore, two linear swimmers in the collinear or side-by-side configuration with the same swimming direction always mutually benefit from passive interactions in the quadrupolar regime (figure \ref{fig_3}, \ref{fig_4}). 

While in some of the previous works passive interaction has been found to be equal to the average local flow field \cite{PooleyAlexanderYeomans2007, AlexanderPooleyYeomans2009} as we do, other authors have reported results for the passive interaction different from the time-averaged flow field. In particular, a quadrupolar passive interaction scaling to leading order as $L^1$ for $L\rightarrow\infty$ has been reported \cite{FarzinRonasiNajafi2012, KurodaYasudaKomura2019}. This is in contrast to the average quadrupolar flow field of a linear-three sphere swimmer which scales as $L^0$ in the same limit \cite{PooleyAlexanderYeomans2007}. 
We have been able to reproduce the linear scaling in $L$ of the passive interaction in a slightly altered perturbative scheme, where the distance between the two swimmers is assumed to be a constantly $r$ throughout one swimming stroke, instead of treating it as an unknown variable to be determined. In this work however, we impose no constraints on the distance between both swimmers, only assuming that the initial distance at $t = 0$ is $r$. Consequently, we obtain the exact correspondence of passive interaction and average unperturbed flow field.

\subsection{Active translation}
Beyond the approximation of the swimmer as a passive object drifting in the local time-averaged flow field around it, the swimmer deforms periodically in order to self-propel, introducing interaction effects which depend on the swimming stroke of both interacting swimmers as well as their relative phase. This effect has been termed active interactions \cite{PooleyAlexanderYeomans2007}.  
In both the force-based and the stroke-based model, active translational interactions arise firstly at the quadrupolar order ($\sim 1/r^3$) and order $A^2$ or $\xi^2$, depending on the choice of model. In both models, this term has only a component parallel to the swimmer axis for arbitrary swimmer positioning. The analytical result for the change in swimming velocity of swimmer $I$ in interaction with a second swimmer $\II$ side-by-side with it reads in the stroke-based model
\begin{eqnarray}
\fl
\Delta v_I^\mathrm{par, SB} = - \frac{a^2 L^2 \omega }{4 r^3 \left(21 a^2-40 a L+16 L^2\right)^2} \times \nonumber \\ \fl \left(12 \left(63 a^2-64 a L+16 L^2\right) (\xi_{I2} \xi_{\II 2} \sin (\beta_I-\beta_{\II}-\delta )+\xi_{I2} \xi_{\II 1} \sin (\beta_I-\delta )+ \right. \label{eq:velQuadrStrokeBased} \\ \fl \left.  \xi_{I1} \xi_{\II 2} \sin (\beta_{\II}+\delta )+\xi_{I1} \xi_{\II 1} \sin (\delta ))+\xi_{\II 1} \xi_{\II 2} \left(441 a^2-792 a L+272 L^2\right) \sin (\beta_{\II})\right), \nonumber
\end{eqnarray}
including both active and passive terms (details of the stroke-based perturbative calculation are presented in \ref{ch:appendixC2}). 
For the force-based model, we restrict the analytical result to two pullers in the same configuration, as otherwise the expression would become too large, 
\begin{eqnarray}
\fl
\Delta v_I^\mathrm{par, FB} = \frac{3 a^2 A_{I 1}^2 \omega }{4 k^2 (7 \nu -4) r^3 \left(16 \Gamma ^2+9 (4-7 \nu )^2\right) \left(16 \Gamma ^2+(4-3 \nu )^2\right)^2} \times \nonumber \\ \fl \left(-32 \Gamma  \left(16 \Gamma ^2 \left(189 \nu ^2-162 \nu +40\right)+3 \left(7371 \nu ^4-21222 \nu ^3+21144 \nu ^2-8992 \nu + \right. \right. \right.  \nonumber \\ \fl \left. \left. \left. 1408\right)\right) \sin (\gamma)+8 \left(512 \Gamma ^4-16 \Gamma ^2 \left(2835 \nu ^3-3870 \nu ^2+1920 \nu -352\right)+9 (9 \nu -4) \right. \right. \label{eq:velQuadrForceBased} \\\fl \left. \left. \left(21 \nu ^2-40 \nu +16\right)^2\right) \cos(\gamma)+(147 \nu -68) \left(256 \Gamma ^4+128 \Gamma^2 \left(9 \nu ^2-18 \nu +8\right)+ \right. \right. \nonumber  \\ \fl \left. \left. 3 (3 \nu -4)^3 (7 \nu -4)\right)\right).\nonumber
\end{eqnarray}
For the details of the force-based calculation we refer to \ref{ch:appendixB}. 
Both results, \eqref{eq:velQuadrStrokeBased} and \eqref{eq:velQuadrForceBased}, scale to leading order constant in $L$ in an expansion around $L = \infty$, similarly to the passive effect \eqref{eq:quadrupolarFlowField}. For moderate stroke amplitudes of both swimmers ($A \approx k a, \xi \approx a$), we therefore observe that passive and active quadrupolar interaction are typically of the same order of magnitude. In the collinear and side-by-side configuration, passive interaction always enhances the self-propulsion of both swimmers, whereas the sign of the active interaction depends on the phase shift between both swimmers via a sine or cosine function, as shown in \eqref{eq:velQuadrStrokeBased}, \eqref{eq:velQuadrForceBased}. Thus, for a large part of the parameter space, two swimmers in such configurations overall benefit in their propulsion (figure \ref{fig_3} (b), \ref{fig_4} (b)), suggesting that a swarm of linear swimmers should propagate faster if the swimmers arrange behind each other or all side-by-side \cite{MirzakhanlooJalaliAlam2018}. However, this result is sensitive to the relative swimmer positioning and orientation. In particular, if the swimmers are arranged on a line or side-by-side, but with opposing swimming directions, the swimmers will predominantly hinder each other's propulsion due to passive effects. 
\begin{figure}
\centering
\includegraphics[scale=0.8]{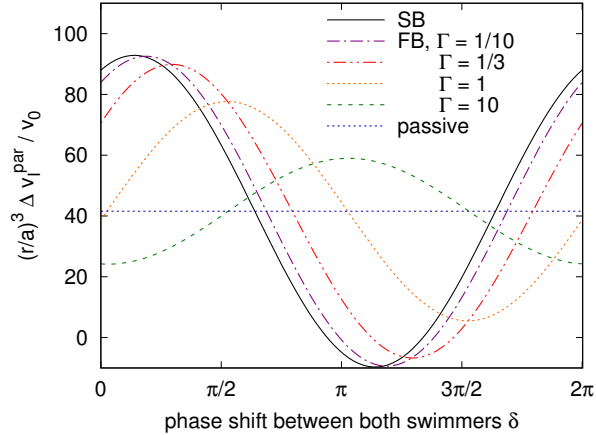}
\caption{Amplitude of the $\xi^2/r^3$ or $A^2/r^3$ translational interaction along the swimmer axis of puller $I$ ($\Delta v^\mathrm{par}_I$), in interaction with puller $\II$ side-by-side to it, in the stroke-based (SB) and the force-based (FB) model for different values of $\Gamma$ in dependence of the phase shift $\delta$ between both swimmers. The horizontal blue line denotes the passive component, which is independent of $\Gamma$ and equal for the SB and FB models. In the force-based model, the driving forces of both swimmers are chosen such that the resulting stroke of a single swimmer equals the stroke employed for the stroke-based swimmers.}
\label{fig_5}
\end{figure}

Comparing the active interactions using the map from force parameters to stroke parameters, we find that the results in both models generally differ, but agree for arbitrary swimmer positioning in the limit $\Gamma = 6 \pi \eta a \omega / k \to 0$, i.e. when the driving frequency in the force-based model $\omega$ becomes small compared to the inverse viscous time $t_V^{-1}$. 
In figure \ref{fig_5}, we plot the total quadrupolar interaction along the swimmer axis to second order in $A$ and $\xi$, respectively, for two side-by-side swimmers in dependence of the phase shift $\gamma$, while varying the rescaled driving frequency $\Gamma$. 
It becomes apparent that for $\Gamma$ approaching zero, the respective force-based curve approaches smoothly the stroke-based curve until they coincide in the limit $\Gamma \to 0$. In contrast, for large values of $\Gamma$, the interaction effect in the force-based model is out of phase by approximately $\pi$ compared to the stroke-based active interaction, which means that while active effects in one model boost the swimmer, they slow it down in the other model. This shows that for frequencies with $\Gamma \approx 1$ or larger, the stroke-based results are insufficient to describe the interaction of elastic swimmers and a theory accounting for the altered swimming stroke is inevitable. While figure \ref{fig_5} corresponds to the interaction of two pullers, qualitatively very similar results are found for the different other combinations of pushers and pullers. Figure \ref{fig_5} is hence representative for interacting linear swimmers, independently of the dipolar swimmer types. 

To compare both models in figure \ref{fig_5}, a certain stroke amplitude and phase between the two swimmer arms had to be assumed for each swimmer in the stroke-based model. The driving forces for the force-based swimmers were then adjusted for each choice of frequency such that at the single swimmer level the resulting stroke amplitude and phase would coincide with the before fixed stroke. We highlight that when comparing the interactions within two-swimmer systems in both approaches, the driving forces on each swimmer have to be adjusted such that they produce the SB model stroke when the swimmer is alone. If we, hypothetically, would adjust the driving forces such that the resulting swimmer strokes would coincide with the SB strokes while the swimmers interact, we would simply recover the stroke-based system since fixed strokes together with the force-free and torque-free conditions already determine the whole system dynamics. 


Analyzing the terms responsible for the active interactions allows to get an intuitive understanding for why they are different in both models. Active quadrupolar translation of swimmer $I$, to second order in $A$ and $\xi$, results from the time-dependent oscillating dipolar flow field, scaling as $A^1$ or $\xi^1$ respectively, produced by swimmer $\II$, interacting with the stroke of swimmer $I$. 
This flow field can be decomposed in a Taylor series around the position of swimmer $I$. The zeroth order, spatially constant term is associated with the same hydrodynamic force on all three beads, an effect averaging out over one stroke cycle due to the purely oscillatory nature of the flow field. 
At the next order in the expansion, the gradient of this instantaneous dipolar flow field, which swimmer $I$ senses along its axis of symmetry, is associated with hydrodynamic forces which expand and compress the swimmer arms and is responsible for the active interaction. 
In the force-based model, the interplay of those forces with the springs and the viscous friction alters the swimming stroke by an additional term $\sim A^1/r^3$. Since the swimmer's velocity is effectively quadratic in its stroke, cross terms of this additional term and the swimmer's original stroke $\sim A^1 r^0$ yield the leading order active interactions with the correct scalings. 
In the stroke-based model, the arm lengths are prescribed a priori such that the gradient of the time-dependent dipolar flow field evokes additional external forces necessary to maintain the prescribed stroke rather than an altered stroke. 
The active interactions then arise as a consequence of the additional forces \cite{FarzinRonasiNajafi2012}.

From this, we can understand why both the stroke-based and force-based models in general differ in the active interactions, but become equivalent when in the force-based model $\Gamma = 6 \pi \eta a \omega / k \to 0$. This limit can also be interpreted as the limit of $k \to \infty$ when fixing $\omega$ and $\eta$, i.e. as the limit of very stiff springs. Consequently, to maintain a constant swimming stroke amplitude of a single swimmer, it is then necessary to scale up the driving force amplitude accordingly. A such force-based swimmer with stiff arms will barely alter its stroke in response to compressing or expanding hydrodynamic forces produced by another nearby swimmer. Instead, the stiff springs will exert forces along the arms counterbalancing the hydrodynamic forces - and the mechanism of the active interactions becomes identical to the one for the stroke-based swimmers. 

Comparing to the literature, we find that the results \eqref{eq:velQuadrStrokeBased} and \eqref{eq:velQuadrForceBased} have not been reported yet. Other authors have obtained results for the quadrupolar active interaction scaling to leading order linear in $L$ \cite{FarzinRonasiNajafi2012, KurodaYasudaKomura2019}. Again, we have reproduced such linear scaling in a calculation employing the altered perturbative scheme, which assumes the distance between the middle beads of both swimmers to be fixed to $r$ throughout the swimming stroke. 
Conversely, we find in our calculation that the active quadrupolar interaction scales to leading order constant in $L$, which is in agreement with our numerical calculations (figure \ref{fig_2}, \ref{fig_3}, \ref{fig_4}) and also backed up by the explanation of the active interaction in terms of time-dependent dipolar flow field, which yields a similar scaling. We are able to confirm that the second order quadrupolar translational interaction acts along the swimmer axis only, a finding which has been reported previously \cite{FarzinRonasiNajafi2012}. 

In contrast to the quadrupolar translation, the octopolar ($\sim 1/r^4$) translational interaction has in general both a component along the swimmer axis, which typically becomes important only at close swimmer separations since it decays quicker than quadrupolar effects, as well as a component acting orthogonal to the swimmer axis. As the quadrupolar translational interaction has no orthogonal component, a transition from the passive dipolar far field interaction to active octopolar interaction at intermediate separations can be observed for $v^\mathrm{ort}$ in the side-by-side swimmer configuration (\ref{fig_4} (d), (e)) at a few tenths of swimmer lengths. For two equal swimmers and $\gamma = 0$, the active orthogonal octopolar component vanishes, therefore the transition is not observed in figure \ref{fig_4} (c) and (f).  
In the stroke-based calculation we obtain for the $\xi^2$ octopolar interaction of two side-by-side swimmers orthogonal to the swimmer axis
\begin{eqnarray}
\fl
v_I^\mathrm{ort, SB} = \frac{9 a L^3 \omega  \left(21 a^2-54 a L+32 L^2\right)}{r^4 (7 a-8 L)^2 (3 a-4 L)} \left( \xi_{I 2} \xi_{\II 2} \sin (\beta_I-\beta_{\II}-\delta )+ \nonumber \right.  \\ \left.  \xi_{I 2} \xi_{\II 1} \sin (\beta_I-\delta )-\xi_{I 1} (\xi_{\II 2} \sin (\beta_{\II} +\delta )+\xi_{\II 1} \sin (\delta )) \right).
\label{eq:swimVelOctoOrth}
\end{eqnarray}
Since all summands in \eqref{eq:swimVelOctoOrth} are dependent on $\delta$, this term is purely active, consistent with the absence of even order $\xi^2$ flow fields and passive interaction for the linear swimmer geometry. This result is, to leading order in $1/L$, in agreement with the corresponding formula from \cite{AlexanderPooleyYeomans2009}. 

Comparing to the corresponding result for the force-based interaction, 
\begin{eqnarray}
\fl
v_I^{\mathrm{ort, FB}} = - \frac{9 a^3 A^2 C  \omega \left(63 \nu ^3-246 \nu ^2+312 \nu -128\right) }{r^4 k^2 (8-7 \nu )^2 \nu ^2 \left(16 \Gamma ^2+(4-3 \nu )^2\right)} \left( -B_I B_{\II} \sin (\alpha_I-\alpha_{\II}-\gamma)+ \nonumber  \right. \\  \left. B_I \sin(\alpha_I-\gamma)-B_{\II} \sin(\alpha_{\II}+\gamma)+\sin(\gamma) \right),
\end{eqnarray}
we find that both results are indeed equivalent with respect to the mapping from the force-based to the stroke-based model. This means, this interaction effect is independent of the spring constant $k$ in the force-based model and equal to the interaction effect for two stroke-based swimmers, assuming that the driving forces one each swimmer correspond to the same single swimmer stroke. This result holds for arbitrary swimmer configurations. In contrast, we find that the component of the octopolar translation along the swimmer axis is not equivalent in both models.

\subsection{Active rotation}
Active rotation is firstly observed at octopolar order ($\sim 1/r^4$) and order $A^2$ and $\xi^2$. In the stroke-based model, we find for two side-by-side swimmers
\begin{eqnarray}
\fl
\Omega_I^\mathrm{SB}  = -\frac{9 a L^2 \omega}{8 r^4 (7 a-8 L) (3a-4 L) (7 a-4 L)^2} \times \nonumber \\ \left(2 a \xi_{\II 1} \xi_{\II 2} \left(1029 a^2-1652 a L+544 L^2\right) \sin (\beta_{\II})  - \left(1617 a^3- \right. \right. \nonumber \\ \left. \left. 3136 a^2 L+1936 a L^2- 384 L^3\right)  (\xi_{I2} \xi_{\II 2} \sin(\beta_I-\beta_{\II}-\delta )+  \right. \\ \left. \xi_{I2} \xi_{\II 1} \sin (\beta_I-\delta ) +\xi_{I1} \xi_{\II 2} \sin (\beta_{\II}+\delta )+ \xi_{I1} \xi_{\II 1} \sin (\delta ))\right), \nonumber
\label{eq:rotationStrokeBased}
\end{eqnarray}
where both the active and passive contribution is included. 
In the force-based model, we again restrict the analytical result to two pullers in the same configuration, finding
\begin{eqnarray}
\fl
\Omega_I^\mathrm{FB} = - \frac{27 a^2 A_{I1}^2 \omega }{4 r^4 k^2 \nu  (7 \nu -8) (7 \nu -4) \left(16 \Gamma ^2+9 (4-7 \nu )^2\right) \left(16 \Gamma ^2+(4-3 \nu )^2\right)} \times \nonumber \\ \left(\left(1617 \nu ^3-3136 \nu ^2+1936 \nu -384\right) \left(16 \Gamma ^2+63 \nu ^2-120 \nu + \right. \right. \nonumber \\ \left. \left. 48\right) \cos (\gamma)- \nu  \left(1029 \nu ^2- 1652 \nu +544\right) \left(16 \Gamma ^2+63 \nu ^2-120 \nu + \right. \right. \\ \left. \left. 48\right)- 8 \Gamma  \left(14553 \nu ^4-34692 \nu ^3+ 29968 \nu ^2- 11200 \nu + 1536\right) \sin (\gamma)\right). \nonumber 
\label{eq:rotationForceBased}
\end{eqnarray}
Employing the mapping from force-based to stroke-based parameters reveals that also this octopolar rotation is equivalent in both models. 
The active octopolar rotation scales to leading order as $L^1$ for $L \rightarrow \infty$, consistent with the results reported in the literature \cite{PooleyAlexanderYeomans2007, AlexanderPooleyYeomans2009, FarzinRonasiNajafi2012}. In particular, we find that our result agrees algebraically in the leading order in $1/L$ with the result reported in \cite{AlexanderPooleyYeomans2009}. 
Unlike the active contribution, the passive contribution to octopolar rotation scales to leading order as $L^0$, as can be seen from \eqref{eq:rotationStrokeBased} and \eqref{eq:rotationForceBased}. Therefore, the active component of the rotation is typically dominating over the passive one,  and clockwise (CW) as well as counterclockwise (CCW) rotation can both be observed when varying $\gamma$ or $\delta$ (figure \ref{fig_4} (b)) due to the sinusoidal dependence of active interaction on the phase shift. 

Investigating the origin of the terms responsible for the active octopolar rotation at this order in the perturbative scheme, we find that three different terms contribute to it.
First, swimmer $I$ rotates with magnitude $\sim A^1/r^4$ in oscillatory fashion due to the curl associated to the time-dependent quadrupolar flow field produced by swimmer $\II$. From the interplay of this time-dependent rotation with the swimmer's own self-propulsion, swimmer $I$ experiences an active overall rotation $\sim A^2/r^4$. 
Second, swimmer $I$ moves within the instantaneous flow field produced by the other swimmer due to its $A^1 r^0$ original swimming stroke, experiencing the curl of this flow field at different positions. The inference of both the oscillating motion and the time-dependent curl of the flow field contributes to the active interactions as well. 
Third, swimmer $I$ experiences hydrodynamic forces which would bend the swimmer, resulting from the instantaneous dipolar flow field produced by swimmer $\II$. The counteracting forces exerted by the rigid joint, although themselves being torque-free, induce fluid flows, which, in interplay with the $A^1 r^0$ swimming stroke of swimmer $I$, give also rise to active rotation. 
All three of these mechanisms are independent of whether the swimmer arms are stiff or elastic, explaining why the active octopolar rotation is equivalent in both models. 

\subsection{Swimmer behavior for arbitrary positioning and long-term behavior}
We now extend our considerations to two arbitrarily positioned force-based swimmers and the temporal evolution of their relative positions. For the latter, different types of long-term trajectories for two swimmers have already been classified in the stroke-based model in dependence on the initial relative positioning \cite{PooleyAlexanderYeomans2007, FarzinRonasiNajafi2012}. Both authors investigating the long-term behavior have obtained broadly similar results for the long-term trajectories, although different driving protocols were employed. 
We extend these studies to force-based swimmers with sinusoidal driving and show that the long term behavior is qualitatively the same as for stroke-based swimmers \cite{PooleyAlexanderYeomans2007, FarzinRonasiNajafi2012} and is also robust with respect to varying the swimmer characteristics between puller- and pusher-type. 
This can be understood from the fact that the long-term behavior is predominantly a result of the self-propulsion and rotational interaction, which are to leading order equivalent in both approaches, and is only weakly influenced by translational interaction \cite{PooleyAlexanderYeomans2007}. In contrast to previous works, we infer the long-term behavior from the instantaneous interactions evaluated in the parameter space of all possible relative swimmer configurations, allowing for a simple graphical representation of the system's dynamics.

\begin{figure}
\centering
\includegraphics[scale=0.805]{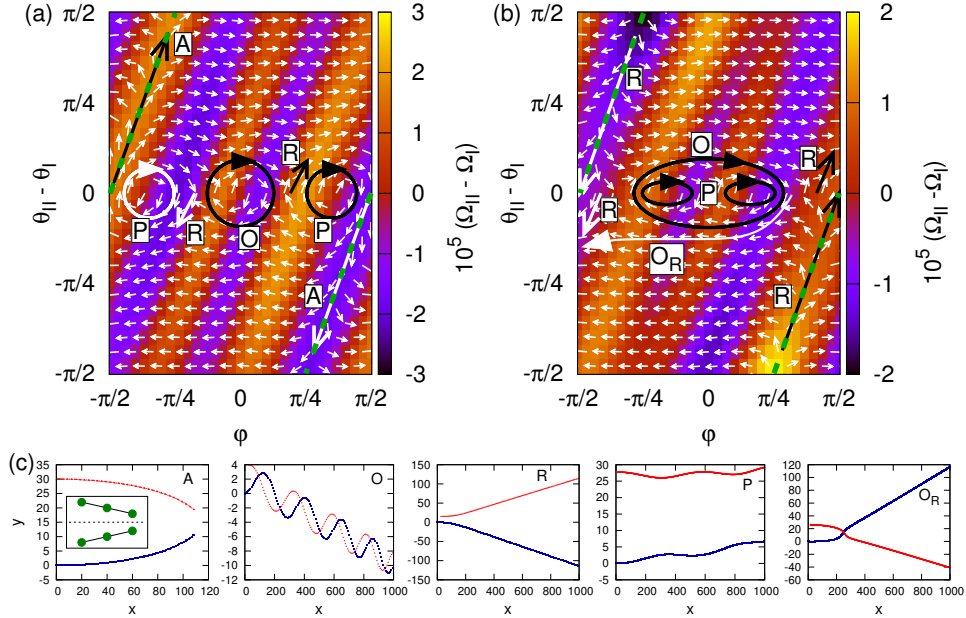}
\caption{Difference of the angular velocities of swimmer $\II$ and $I$ in dependence of angles encoding the relative swimmer positions, $\varphi$ and $\theta_2 - \theta_1$, for two force-based pullers with (a) phase shift $\gamma = 0$ and (b) $\gamma = \pi$ for distance $r = 30 a$. The small white arrows indicate the time evolution that the system undergoes. The length of the arrows has been normalized for better visibility. The large black and white arrows illustrate the different types of long-term behavior observed, the different colors are to improve the contrast of the figure in print. 
In (c), typical typical trajectories of the two swimmers moving from left to right are depicted. Point of reference is the central bead in each swimmer.}
\label{fig_6}
\end{figure}

The relative positioning of both swimmers is parametrized by their distance $r$, the orientation of the connection between the middle beads of both swimmers relative to the axis of swimmer $I$, $\varphi$, and the difference between the two swimmer orientations,  $\theta_{\II} - \theta_I$ (figure \ref{fig_1}). 
The equations governing the momentary evolution of a system of two swimmers, obtained by elementary geometry, are given by 
\begin{align}
&\frac{d}{dt} \varphi = \frac{v_{0, I} + \Delta v_{I}^\mathrm{par}}{r} \sin \varphi - \frac{v_I^\mathrm{ort}}{r} \cos \varphi - \frac{v_{0, \II} + \Delta v_{\II}^\mathrm{par}}{r} \sin (\varphi^{\prime}) - \frac{v_{\II}^\mathrm{ort}}{r} \cos(\varphi^{\prime})   - \Omega_I, \label{eq:phiDot} \\
&\frac{d}{dt} (\theta_{\II} - \theta_I) = \Omega_{\II} - \Omega_I, \label{eq:thetasDot} \\
&\frac{d}{dt} r =- (v_{0, I} + \Delta v_{I}^\mathrm{par}) \cos \varphi - v_I^\mathrm{ort} \sin \varphi - (v_{0, \II} + \Delta v_{\II}^\mathrm{par}) \cos (\varphi^{\prime}) + v_{\II}^\mathrm{ort} \sin(\varphi^{\prime}), 
\end{align}
with $\varphi^\prime = \pi - \varphi + \theta_{\II} - \theta_I$. $\Delta v_s^\mathrm{par}$, $v_{s}^\mathrm{ort}$ and $\Omega_s$ denote the time-averaged interaction effect parallel to the swimmer axis, orthogonal to the swimmer axis and the rotational interaction effect on swimmer $s$. $v_{0, s}$ denotes the corresponding time-averaged single swimmer speed. 
The first four terms on the right hand side of \eqref{eq:phiDot} correspond to how $\varphi$ changes due to the swimmer translation, while the last term accounts for the effect due to the rotation of swimmer $I$. To illustrate the behavior of both swimmers, we display the difference in the angular velocities of both swimmers, $\Omega_{\II} - \Omega_I$, time-averaged over a complete stroke cycle in dependence of $\varphi$ and $\theta_{\II} - \theta_I$ for a fixed distance $r$ by the color in figure \ref{fig_6} (a) and (b). The momentary time evolution of the system is illustrated by arrows corresponding to the vector field
\begin{equation}
X_t = \left(\frac{d}{dt} \varphi, \frac{d}{dt} (\theta_{\II} - \theta_I) \right). 
\label{eq:temporalVectorField}
\end{equation}
An arrow pointing to the left or right indicates that $\varphi$ increases or decreases, and an arrow pointing upwards or downwards corresponds to $\theta_{\II} - \theta_I$ increasing or decreasing. When both $\varphi$ and $\theta_{\II} - \theta_I$ change with time, the arrow is tilted such that the horizontal and the vertical component represent the corresponding rate of change. For better visibility the lengths of all vectors have been normalized. We obtain qualitatively similar plots for two interacting pushers or for combinations of a pusher and a puller. Figure \ref{fig_6} can therefore be considered representative for the interaction of linear swimmers at intermediate distances up hundred swimmer lengths, independently of the dipolar characteristics.
This effectively results from the fact that the linear swimmer with equal bead radii produces only a weak average dipolar flow field, as discussed before, and interaction at the distance of $r = 30 a$ considered here is dominated by quadrupolar effects. It is in contrast to swimmer models with dominant dipolar average flow fields, which often exhibit strong differences in the collective behavior of pullers and pushers \cite{Hoell2018, Yoshinaga2017, Bardfalvy2020}. 

The time evolution for two in-phase swimmers ($\gamma = 0$, figure \ref{fig_6} (a)) and two swimmers out-of-phase by $\gamma = \pi$ (figure \ref{fig_6} (b)) is dominated for a major part of the parameter space spanned by $\varphi$ and $(\theta_{\II} - \theta_I)$ by the self-propulsion terms in \eqref{eq:phiDot}, which are linear in $v_{0, s}$. In these parts of the parameter space, we have $|d/dt \, \varphi| \gg |d/dt \, (\theta_{\II} - \theta_I)|$, associated with horizontal arrows. Exceptions from this are first the regions around the green dashed lines, corresponding to $\theta_{\II} - \theta_I = 2 \varphi \pm \pi$, and second the region around the $\varphi$-axis, i.e. $\theta_{\II} - \theta_I = 0$, corresponding to swimmers with parallel swimming direction. It is straight forward to verify that the first case corresponds to both swimmers being symmetric with respect to some axis in the $x$-$y$-plane, as sketched in the inset in the first panel of figure \ref{fig_6} (c). In both of the two cases, the self-propulsion terms proportional to $v_{0, s}$ in \eqref{eq:phiDot} cancel. Therefore, the swimmer behavior in these regions is dominated by rotational interactions. In the symmetric configuration (green dashed lines), in-phase pullers rotate towards each other (which can be seen from the vectors pointing down at $\varphi = \pi/2, \theta_{\II} - \theta_{I} = 0$ in \ref{fig_6} (a)), but rotate away from each other when out-of-phase, consistently with figure \ref{fig_4} (b).

During swimmer interaction, not only the values of $\varphi, \theta_{I}$ and $\theta_{\II}$ change, but also the distance $r$ between both swimmers does. 
Varying the value of $r$, the angular velocities $\Omega_I$ and $\Omega_{\II}$ scale as $r^{-4}$ as long as $r$ does not leave the regime dominated by the quadrupolar flow fields. Hence, the color structure in figure \ref{fig_6} stays constant while only the value corresponding to each color is rescaled. 
Conversely, the arrows align with the horizontal direction when $r$ is increased, since the horizontal component \eqref{eq:phiDot} scales to leading order as $r^{-1}$ due to $v_{0, s}$ being independent of $r$. In contrast, the vertical component \eqref{eq:thetasDot} scales as $r^{-4}$. 

Although figure \ref{fig_6} (a) and (b) correspond to constant $r$ and cannot account for the swimmers coming closer or separating, we are able to reproduce the different types of long-term trajectories (figure \ref{fig_6} (c)), which have been similarly reported for two interacting linear stroke-based swimmers with initially parallel swimming direction (figure 4 (a) and (b) in \cite{PooleyAlexanderYeomans2007} and figure 3, upper left panel in \cite{FarzinRonasiNajafi2012}). In figure \ref{fig_6} (c), the middle bead of each swimmer is taken as reference for the swimmer position. Since the translational interaction is small compared to self-propulsion, both swimmers are in general oriented tangential to their trajectories, swimming from left to right.

We first consider two swimmers with zero phase shift $\gamma$ (figure \ref{fig_6} (a)) and classify the different types of long-term trajectories observed when the swimmers have initially parallel swimming direction ($\theta_{\II} - \theta_I = 0$) but are positioned arbitrarily otherwise, corresponding to varying the initial value of $\varphi$. 
Two approximately collinear swimmers, i.e. $\varphi \approx 0$, swim on oscillatory (O) trajectories. Mapping this behavior to the phase space of $\varphi$ and $\theta_{\II} - \theta_I$, it corresponds to the curl around $\varphi = 0, \theta_{\II} - \theta_I = 0$ found in figure \ref{fig_6} (a). 
With increasing $\varphi$, we next enter the repulsive regime (R) in which the swimmers rotate away from each other and thus separate such that rotational interaction dies out and the swimmers self-propel on straight trajectories away from each other. Increasing $\varphi$ further, still keeping the swimmers initially parallel, the swimmers typically move on parallel trajectories (P), associated to the curl close to $\varphi \approx \pi/3, \theta_{\II} - \theta_I = 0$. Since this curl is associated to positive values of $\varphi$ only, the swimmer trajectories do not cross, distinguishing it from the oscillatory case. 
For $\varphi$ close to $\pi/2$, the swimmers rotate towards each other, effectively attracting each other (A), until the beads come so close that the Oseen approximation is not valid anymore. This behavior corresponds to the area around the green dashed lines in figure \ref{fig_6} (a). 

Similarly, also the types of trajectories of two out-of-phase swimmers are accounted for in the respective plot of the $\varphi$-$(\theta_{\II} - \theta_I)$ space (figure \ref{fig_6} (b)). Besides the types of trajectories already mentioned, we also observe trajectories where the swimmers first rotate towards each other, their trajectories cross and the swimmers subsequently repel ($\mathrm{O}_\mathrm{R}$). Also for this behavior, the qualitative trajectory in terms of $\varphi$ and $\theta_{\II} - \theta_I$ can be depicted in figure \ref{fig_6} (b).

\section{Summary and Conclusions}
\label{sec:conclusion}
We have calculated the scaling function, the strength, and the direction of the hydrodynamic interactions between two linear three-bead swimmers. Both the force-based and stroke-based models were considered within perturbative approaches, and the results were compared to direct numerical results with excellent agreement. This framework is applicable to arbitrary positioning of the two interacting swimmers, however, the most simple cases of collinear and side-by-side swimmers have been used to illustrate the main results of the paper. 

In both the force-based and the stroke-based model, the swimmers experience passive hydrodynamic interactions given by the time-averaged flow field produced by all nearby swimmers. Comparing the interaction of force-based and stroke-based swimmers under the constraint that the individual devices display identical actuation, passive interactions are equivalent in both models.

In terms of active hydrodynamic interactions, which are a result of the interference of the time-dependent swimming strokes of both swimmers, active rotation and translational interaction orthogonal to the swimmer axis are equivalent to the leading order in the stroke- and the force-based model. However, the leading order of the active translation along the swimmer axis is not. This shows that the interaction is in general not conserved with respect to the mapping between stroke-based and force-based swimmers and that the deformation of one swimmer in the presence of others significantly alters the change in propulsion speed that a swimmer experiences, depending on the design. However, we have shown that both models are equivalent when the driving frequency is small compared to the inverse viscous time, which is meaningful only in the force-based model. Consequently, the stroke-based interaction should be regarded as the special case of the force-based interaction in the limit of small frequencies. 

At the leading order in the swimmer actuation, the time-averaged flow field produced by a three-sphere swimmer is proportional to its velocity, with a proportionality coefficient depending solely on the swimmer geometry. This novel result, valid for force-based as well as stroke-based swimmers, allows us to infer the leading order passive interactions directly from the geometry of a system of two interacting swimmers. Applied to the three-bead swimmers this has interesting consequences. Namely, compared to the velocity of a single swimmer, passive interactions enhance the swimming velocity if swimmers are one behind another, or side-by-side in the same direction. This cooperative effect for pairs of swimmers is independent of the details of the driving and is found similarly for pullers and pushers. It relies on the passive quadrupolar interactions, which dominate in the major part of the parameter space. 

In the long-time limit, the behavior of a force-based two-swimmer system is independent of the pusher/puller characteristics of both swimmers up to separations of hundreds of swimmer lengths. We find that the long-term behaviors for force-based and stroke-based swimmers agree qualitatively, which is a result of the equivalence of the leading order rotational interactions in both models. Following a numerical analysis, we have also shown that the different types of long-term trajectories can be qualitatively inferred directly from the instantaneous behavior of both swimmers evaluated in the space of all relative swimmer configurations. 

The insights presented here may help to better understand the collective motion of swarms. In particular, we have thoroughly studied active interactions that rely on  resolving the flow fields produced by each swimmer on time scales below the single stroke. This feature is typically not taken into account in studies of swimmer suspensions, and given that active contribution may be important, accounting for these effects may give rise to new insights in the collective behavior. Our analytical approach may be a particularly suitable tool for the task, as it can be applied explicitly to an arbitrary number of swimmers, the limiting factor being only the tractability. Furthermore, the approach can be easily tuned in terms of the precision of the hydrodynamic interactions at smaller bead separations. This can be done by using the Rotne-Prager approximation or some higher order theory instead of the Oseen approximation, a feature which becomes necessary when the density of swimmers increases. Given that hydrodynamic interactions have been shown to be important even in dense suspensions of bacteria \cite{IshikawaHota2006}, the emerging results may find application well beyond the physical context presented herein.

\section*{Acknowledgements}
We thank the Priority Programme 'Microswimmers - From Single Particle Motion to Collective behavior' (SPP 1726) of the Germany Research Foundation for funding, and acknowledge the support of the Excellence Cluster 'Engineering of Advanced Materials' at the Friedrich-Alexander-University Erlangen-Nuremberg. We benefited from insightful discussions with Alexander Sukhov, Oleg Trosman and Nicolas Vandewalle.

\appendix
\section{Rigidity constraint in the force-based numerical calculation}
\label{ch:appendixA}
\setcounter{section}{1}
To restore a swimmer's linear shape while interacting with other swimmers, we impose in the numerical calculations an additional harmonic angle potential at the middle bead of each swimmer. 
We prescribe it as 
\begin{equation}
    \phi_s = k_A \, k \, a^2 \, \kappa_s^2,
\end{equation}
with $\kappa_s$ the angle enclosed by $\vec{R}_{s2} - \vec{R}_{s1}$ and $\vec{R}_{s3} - \vec{R}_{s2}$, i.e. $\kappa_s = 0$ corresponds to straight shape, and $k_A$ the relative stiffness of the angular spring. A value of $k_A = 100$ typically suffices to suppress bending, but is small enough to not make the system stiff. The forces resulting from the angle potential satisfy the force- and torque-free condition.

\section{Details on the force-based perturbative calculation}
\label{ch:appendixB}
The perturbative scheme employed in this paper is derived from the calculation scheme explained in detail in \cite{ZieglerHubertVandewalle2019}, with the only differences that firstly two interacting swimmers are considered instead of one, secondly forces preventing the swimmers from bending have to be imposed and thirdly an additional perturbation in $q$ is performed.
  
The equation of motion \eqref{eq:EOM} with additional anti-bending forces can be cast into the form 
\begin{equation}
\frac{d}{d \tau} \bm{R}' = \underline{\mu}' (\bm{R}') \left[ \epsilon \bm{E}' (\bm{R}', \tau) + \left( \epsilon \bm{E}_A^{\prime (1)} (\bm{R}', \tau) + \epsilon^2 \bm{E}_A^{\prime (2)} (\bm{R}', \tau) + ... \right) + \bm{G}'(\bm{R}') \right].
\label{eq:EOMCompact}
\end{equation}
All dashed variables denote rescaled quantities with respect to the bead radius $a$ and the viscous time $t_V$ \cite{ZieglerHubertVandewalle2019}, $\tau := t/t_V$ denotes the rescaled time $t$ and $\epsilon := A/(ka)$. In particular, we have introduced $(n \cdot d)$-dimensional bold vectors for the rescaled position
\begin{equation}
\bm{R}' := \frac{1}{a} (\vec{R}_{I 1}, \vec{R}_{I 2},  \vec{R}_{I 3}, \vec{R}_{\II 1}, \vec{R}_{\II 2},  \vec{R}_{\II 3})
\end{equation}
and the external driving forces
\begin{equation}
\bm{E}' (\bm{R}', \tau) := \frac{1}{A} (\vec{E}_{I 1} (\tau t_V), ..., \vec{E}_{\II 3} (\tau t_V)).
\end{equation}
The driving forces as defined in \eqref{eq:drivingForces} have to be treated in the perturbative formalism as dependent on the bead positions because they act along the swimmer orientation vectors $\vec{n}_s$. 
Denoting the total spring force on each bead $sj$ by 
\begin{equation}
\vec{G}_{sj} (\vec{R}_{I 1}, ..., \vec{R}_{\II 3}) := \sum_{m \in \mathrm{NN}(j)} \vec{g}(\vec{R}_{sm} - \vec{R}_{sj}),
\end{equation}
we define 
\begin{equation}
\bm{G}'(\bm{R}') := \frac{1}{ka} \left( \vec{G}_{I1} (a \bm{R}'), ..., \vec{G}_{\II 3} (a \bm{R}') \right),
\end{equation}
as the vector of all rescaled spring forces on all the beads. 
The mobility matrix $\underline{\mu}' (\bm{R}')$ with $(n \cdot d) \times (n \cdot d)$ components is defined by $n \times n$ blocks with in the diagonal blocks $\hat{1}/(6 \pi \eta a)$ the self-mobilities and in all other blocks the corresponding Oseen tensor (see (8) in \cite{ZieglerHubertVandewalle2019}). 

In addition to the driving forces, which act along the respective swimmer axis in the form specified in \eqref{eq:drivingForces}, we include 'anti-bending' forces for each swimmer and at each order in $\epsilon$. They are denoted, in rescaled form, by $\bm{E}_A^{\prime (i)} (\bm{R}', \tau)$ for forces at order $\epsilon^i$. 
The form of the anti-bending forces on each bead of a swimmer is derived from a potential defined to be the angle enclosed between both swimmer arms, 
\begin{equation}
    \Phi_s = \kappa_s.
\end{equation}
In contrast to the potential $\phi_s$ defined in \ref{ch:appendixA}, which attains a minimum at $\kappa_s = 0$ corresponding to zero anti-bending forces, the potential $\Phi_s$ is associated with non-zero forces that would bend the swimmer $s$ even if it has linear shape. 
We then define the anti-bending forces as
\begin{equation}
    \bm{E}_A^{\prime (i)} (\bm{R}^{\prime}, \tau) =  A^{\mathrm{bend}, (i)}_{I} (\tau)  \nabla_{\bm{R}'} \Phi_I + A^{\mathrm{bend}, (i)}_{\II} (\tau)  \nabla_{\bm{R}'} \Phi_{\II},
\end{equation}
with $A^{\mathrm{bend}, (i)}_{I} (\tau)$, $A^{\mathrm{bend}, (i)}_{\II} (\tau)$ a priori undetermined time-dependent prefactors. This definition ensures that the overall force and torque exerted by the anti-bending forces on each swimmer vanish. 

At order $\epsilon^1$ we typically observe bead oscillations with the base frequency \cite{ZieglerHubertVandewalle2019}, hence it suffices to assume for the explicit time dependence
\begin{equation}
A^{\mathrm{bend}, (1)}_s (\tau) := A^{\mathrm{bend}, (1), \mathrm{sin \, 1}}_s \sin(\Gamma \tau) +  A^{\mathrm{bend}, (1), \mathrm{cos \, 1}}_s \cos(\Gamma \tau). 
\end{equation}
At order $\epsilon^2$, it is necessary to include oscillating terms with the second harmonic and a constant component in the anti-bending forces: 
\begin{equation}
A^{\mathrm{bend}, (2)}_s (\tau) := A^{\mathrm{bend}, (2), \mathrm{sin \, 2}}_s \sin(2 \Gamma \tau) +  A^{\mathrm{bend}, (2), \mathrm{cos \, 2}}_s \cos(2 \Gamma \tau) + A^{\mathrm{bend}, (2), \mathrm{const}}_s. 
\end{equation}
The prefactors will be carried along in the calculation of the swimmer behavior at each order $\epsilon^i$ and then determined a posteriori such that they effectively cancel out any bending of the swimmers. 

To solve \eqref{eq:EOMCompact} by perturbation theory around $\epsilon = 0$ and $q = 0$, corresponding to zero driving forces and infinitely separated swimmers, we introduce the displacement of all beads with respect to their initial position, $\bm{\xi}' (t) := \bm{R}' (t) - \bm{R}^{\prime \mathrm{eq}}$, following the notation in \cite{ZieglerHubertVandewalle2019}, with $\bm{R}^{\prime \mathrm{eq}}$ the vector of all rescaled initial bead positions. We assume that in this configurations both swimmers are in mechanical equilibrium, i.e. the distance between neighboring beads is $L$. We then expand the displacement as a power series in both $\epsilon$ and $q$,
\begin{equation}
    \bm{\xi}' (\tau) = \sum_{i = 1}^\infty \sum_{j = 0}^\infty \epsilon^i q^j \bm{\xi}^{\prime (i, j)} (\tau).
    \label{eq:expansionXi}
\end{equation}
The first index of $\bm{\xi}^{\prime}$ is associated to powers of $\epsilon$ and the second index is associated to powers of $q$. 
For simplicity, we subsequently omit the time-dependence of $\bm{\xi}'$ and $\bm{R}'$. 

We proceed by expanding $\underline{\mu}' (\bm{R}')$, $\bm{G}'(\bm{R}')$ as well driving and anti-bending forces in a Taylor series around the equilibrium configuration of both swimmers, $\bm{R}^{\prime \mathrm{eq}}$, and insert the expansion \eqref{eq:expansionXi} for the bead displacement (see \cite{ZieglerHubertVandewalle2019} for details of the expansion). 
At each order of $\epsilon^i q^j$ the equation of motion can then be cast into the form
\begin{equation}
\frac{d}{d \tau} \bm{\xi}^{\prime (i, j)} = \underline{K}^{\prime (0, 0)} \bm{\xi}^{\prime (i, j)} + \bm{S}^{\prime (i, j)} (\tau).
\label{eq:EOMSplitByEpsAndQ}
\end{equation}
Here, $\underline{K}^{\prime (0, 0)}$ denotes the $q^0$ component of the $\epsilon$-independent $\underline{K}'$ defined by
\begin{equation}
    \underline{K}' := \underline{\mu}'(\bm{R}^{\prime \mathrm{eq}}) \cdot \nabla_{\bm{R}'} \bm{G}' (\bm{R}^{\prime \mathrm{eq}}),
\end{equation}
with $\nabla_{\bm{R}'}$ the gradient with respect to all $(n \cdot d)$ bead positions, and $\bm{S}^{\prime (i, j)} (\tau)$ a source term which pools all terms independent of $\bm{\xi}^{\prime (i, j)}$. Since the expansion of the bead displacements \eqref{eq:expansionXi} contains only terms with positive exponents in $\epsilon$ and $q$, the source term $\bm{S}^{\prime (i, j)} (\tau)$ can only contain components of the displacement $\bm{\xi}^{\prime (i', j')}$ with $i' \leq i, j' \leq j$, but not $i' = i$ and $j' = j$. 

The source term $\bm{S}^{\prime (i, j)} (\tau)$ composes at each order $\epsilon^i q^j$ of a contribution resulting from the expansion in $\epsilon$ and a contribution resulting from the expansion in $q$, $\bm{S}^{\prime (i, j)} (\tau) = \bm{S}^{\prime (i, j)}_\epsilon (\tau) + \bm{S}^{\prime (i, j)}_q (\tau)$.
The first contribution is given similarly as in the case of a single swimmer (equations (21) and (22) in \cite{ZieglerHubertVandewalle2019}) for the first two orders in $\epsilon$ as, 
\begin{equation}
    \bm{S}^{\prime (1, \mathrm{all})}_\epsilon = \underline{\mu}' (\bm{R}^{\prime \mathrm{eq}}) \left( \bm{E}' (\bm{R}^{\prime \mathrm{eq}}, \tau) +  \bm{E}^{\prime (1)}_A (\tau) \right),
\end{equation}
\begin{eqnarray}
\fl
    \bm{S}^{\prime (2, \mathrm{all})}_\epsilon = \underline{\mu}' (\bm{R}^{\prime \mathrm{eq}}) \left( \frac{1}{2} \bm{\xi}^{\prime (1)} \cdot (\bm{\xi}^{\prime (1)} \cdot (\nabla_{\bm{R}'} \nabla_{\bm{R}'} \bm{G}') (\bm{R}^{\prime \mathrm{eq}})) + \bm{\xi}^{\prime (1)}\cdot \nabla_{\bm{R}'} \bm{E}' (\bm{R}^{\prime \mathrm{eq}}, \tau) + \right. \nonumber \\ \fl \left. \bm{E}^{\prime (2)}_A (\tau) \right) + \left( \bm{\xi}^{\prime (1)} \cdot \nabla_{\bm{R}'} \underline{\mu}' (\bm{R}^{\prime \mathrm{eq}}) \right) \left( \bm{E}' (\bm{R}^{\prime \mathrm{eq}}, \tau) +  \bm{E}^{\prime (1)}_A (\tau) + \bm{\xi}^{\prime (1)} \cdot \nabla_{\bm{R}'} \bm{G}' (\bm{R}^{\prime \mathrm{eq}}) \right) \nonumber \\
\end{eqnarray}
Here we use the notation $\bm{S}^{\prime (i, \mathrm{all})} = \sum_j q^j \bm{S}^{\prime (i, j)}$.  
The expressions for $\bm{S}^{\prime (i, j)}_\epsilon (\tau)$ are then obtained by expanding the above expressions by powers of $q$. 
The second contribution $\bm{S}^{\prime (i, j)}_q$, due to the application of perturbation theory in $q$, is given at each order $\epsilon^i q^j$ by
\begin{equation}
    \bm{S}^{\prime (i, j)}_q = \sum_{l = 1}^j \underline{K}^{\prime (0, l)} \bm{\xi}^{\prime (i, j - l)}.
\end{equation}

Since $\bm{S}^{\prime (i, j)} (\tau)$ depends only on the components of the displacement $\bm{\xi}^{\prime (i', j')}$ with $i' \leq i, j' \leq j$ but not $i' = i$ and $j' = j$, an ascending iteration scheme through the orders in $\epsilon$ and $q$ allows to directly calculate at each order first the source term and second from this the bead displacement by solving \eqref{eq:EOMSplitByEpsAndQ}. Being interested in the swimmer behavior up to e.g. order $\epsilon^2 q^4$, a suitable iteration scheme is given by  $\epsilon^1 q^0 \to \epsilon^1 q^1 \to \epsilon^1 q^2 \to \epsilon^1 q^3 \to \epsilon^1 q^4 \to \epsilon^2 q^0 \to \epsilon^2 q^1 \to \epsilon^2 q^2 \to \epsilon^2 q^3 \to \epsilon^2 q^4$. 

To solve \eqref{eq:EOMSplitByEpsAndQ} at each order, we decompose the respective source term into an oscillating component and a constant component,
\begin{equation}
    \bm{S}^{\prime (i,j)} = \sum_f \bm{S}^{\prime (i,j), \mathrm{sin\, f}} \sin(f \Gamma \tau) + \sum_f \bm{S}^{\prime (i,j), \mathrm{cos\, f}} \cos(f \Gamma \tau) + \bm{S}^{\prime (i,j), \mathrm{const}}, 
\end{equation}
with $f \leq i$ an integer and $\bm{S}^{\prime (i,j), \mathrm{sin \, f}}, \bm{S}^{\prime (i,j), \mathrm{cos \, f}}, \bm{S}^{\prime (i,j), \mathrm{const}}$ prefactors to be determined by decomposition. 
Due to the linearity of \eqref{eq:EOMSplitByEpsAndQ}, one can calculate the solution for each component separately and superimpose those solutions to obtain the final result for the bead displacement. 
For the oscillating source terms, the corresponding solution is given by \cite{Felderhof2006, ZieglerHubertVandewalle2019}
\begin{align}
\bm{\xi}^{\prime (i, j)} = & \sum_f \left(f^2 \Gamma^2 \underline{1} + \underline{K}^{\prime (0, 0)} \right)^{-1} \left( f \Gamma \bm{S}^{\prime (i,j), \mathrm{cos \, f}} - \underline{K}^{\prime (0, 0)} \bm{S}^{\prime (i,j), \mathrm{sin \, f}}  \right) \sin(f \Gamma \tau)  \\ - & \sum_f \left(f^2 \Gamma^2 \underline{1} + \underline{K}^{\prime (0, 0)} \right)^{-1} \left( f \Gamma \bm{S}^{\prime (i,j), \mathrm{sin \, f}} + \underline{K}^{\prime (0, 0)} \bm{S}^{\prime (i,j), \mathrm{cos \, f}} \right) \cos(f \Gamma \tau).
\end{align}

The constant contribution to the source term gives rise to both the translational and rotational velocities of both swimmers as well as to deformations. 
Solving \eqref{eq:EOMSplitByEpsAndQ} becomes simple by decomposing the source term in the eigenbasis of $\underline{K}^{\prime (0, 0)}$. By its definition, $\underline{K}^{\prime (0, 0)}$ is independent of $q$ and thus does not couple between both swimmers. As an example, for $\theta_I = \theta_{\II} = 0$ the matrix $\underline{K}^{\prime (0, 0)}$ is given by
\begin{eqnarray}
\fl
    \underline{K}^{\prime (0, 0)} = \nonumber \\
    \fl
    \begin{pmatrix}
    \frac{3}{2} \nu -1 & 0 & 1 - \frac{9}{4} \nu & 0 & \frac{3}{4} \nu & 0 & 0 & 0 & 0 & 0 & 0 & 0 \\
    0 & 0 & 0 & 0 & 0 & 0 & 0 & 0 & 0 & 0 & 0 & 0\\
    1 - \frac{3}{2} \nu & 0 & 3 \nu -2 & 0 & 1 - \frac{3}{2} \nu & 0 & 0 & 0 & 0 & 0 & 0 & 0 \\
    0 & 0 & 0 & 0 & 0 & 0 & 0 & 0 & 0 & 0 & 0 & 0\\
    \frac{3}{4} \nu & 0 & 1 - \frac{9}{4} \nu & 0 &  \frac{3}{2} \nu -1 & 0 & 0 & 0 & 0 & 0 & 0 & 0 \\
    0 & 0 & 0 & 0 & 0 & 0 & 0 & 0 & 0 & 0 & 0 & 0\\
    0 & 0 & 0 & 0 & 0 & 0 & \frac{3}{2} \nu -1 & 0 & 1 - \frac{9}{4} \nu & 0 & \frac{3}{4} \nu & 0 \\
    0 & 0 & 0 & 0 & 0 & 0 & 0 & 0 & 0 & 0 & 0 & 0\\
    0 & 0 & 0 & 0 & 0 & 0 & 1 - \frac{3}{2} \nu & 0 & 3 \nu -2 & 0 & 1 - \frac{3}{2} \nu & 0 \\
    0 & 0 & 0 & 0 & 0 & 0 & 0 & 0 & 0 & 0 & 0 & 0\\
    0 & 0 & 0 & 0 & 0 & 0 & \frac{3}{4} \nu & 0 & 1 - \frac{9}{4} \nu & 0 &  \frac{3}{2} \nu -1 & 0\\
    0 & 0 & 0 & 0 & 0 & 0 & 0 & 0 & 0 & 0 & 0 & 0\\
    \end{pmatrix}.
\end{eqnarray}
Similarly to the case of a single swimmer \cite{ZieglerHubertVandewalle2019}, the components of the constant source term associated with eigenvalue zero in the eigensystem of $\underline{K}^{\prime (0, 0)}$ correspond to translation or rotation of one of the swimmers. In contrast, the components associated with negative eigenvalues correspond to swimmer deformation along its axis, i.e. the average arm lengths of the swimmer differ from the spring length in the mechanical equilibrium. 
Positive eigenvalues are excluded due to the stability of the mechanical equilibrium of the swimmer \cite{ZieglerHubertVandewalle2019}. 

In the eigenbasis of $\underline{K}^{\prime (0, 0)}$, \eqref{eq:EOMSplitByEpsAndQ} assumes for the constant source term the form 
\begin{equation}
    \frac{d}{d \tau} \left( \overline{\bm{\xi}}^{\prime (i, j)} \right)^\delta = \lambda_{(\delta)} \left( \overline{\bm{\xi}}^{\prime (i, j)} \right)^\delta + \left( \overline{\bm{S}}^{\prime (i, j), \mathrm{const}} \right)^{\delta},
\end{equation}
with $\lambda_{(\delta)}, \delta = 1, ..., n \cdot d$ the eigenvalues of $\underline{K}^{\prime (0, 0)}$. The overline denotes expression of a vector with respect to the eigenbasis of $\underline{K}^{\prime (0, 0)}$ and the additional upper index $\delta$ the respective component of the vector. The translational and rotational velocities of each swimmer are thus given directly by the components of the constant source term corresponding to zero eigenvalues, since in this case $\lambda_{(\delta)} = 0$. In order to determine these components, we use a change of basis from the eigensystem of $\underline{K}^{\prime (0, 0)}$ to standard coordinates, $\underline{M}'$, given by 
\begin{equation}
    \underline{M}' = 
    \begin{pmatrix}
    0 & 0 & 0 & 0 & 0 & 0 & 0 & 0 & 0 & 0 & 0 & 1\\
    0 & 0 & 0 & 0 & 0 & 0 & 1 & 0 & 1 & 0 & 1 & 0\\
    0 & 0 & 0 & 0 & 0 & 0 & 0 & 0 & 0 & 1 & 0 & 0\\
    0 & 0 & 0 & 0 & 0 & 0 & 0 & 1 & 0 & 0 & 0 & 0\\
    0 & 0 & 0 & 0 & 0 & 1 & 0 & 0 & 0 & 0 & 0 & 0\\
    1 & 0 & 1 & 0 & 1 & 0 & 0 & 0 & 0 & 0 & 0 & 0\\
    0 & 0 & 0 & 1 & 0 & 0 & 0 & 0 & 0 & 0 & 0 & 0\\
    0 & 1 & 0 & 0 & 0 & 0 & 0 & 0 & 0 & 0 & 0 & 0\\
    0 & 0 & 0 & 0 & 0 & 0 & -1 & 0 & 0 & 0 & 1 & 0\\
    -1 & 0 & 0 & 0 & 1 & 0 & 0 & 0 & 0 & 0 & 0 & 0\\
    0 & 0 & 0 & 0 & 0 & 0 & 1 & 0 & -2 + \frac{6 \nu}{9 \nu -4} & 0 & 1 & 0\\
    1 & 0 & -2 + \frac{6 \nu}{9 \nu -4} & 0 & 1 & 0 & 0 & 0 & 0 & 0 & 0 & 0\\
    \end{pmatrix}.
\end{equation}
With $\underline{Q}' = \mathrm{Diag}(1, 1, 1, 1, 1, 1, 1, 1, 0, 0, 0, 0)$ the diagonal matrix with unity in the first eight entries and zeros everywhere else, translations and rotations of both swimmers are obtained as
\begin{equation}
    \bm{u}' = \underline{M}' \cdot \underline{Q}' \cdot \underline{M}^{\prime -1} \bm{S}^{\prime (i, j)}_\mathrm{const}.
\end{equation}
A suitable decomposition of these bead velocities into translation along the swimmer axis, orthogonal to it and rotation of each swimmer yields the respective velocities. Taylor expanding these velocities by powers of $q$ then allows to identify quadrupolar, octopolar etc interaction effects. 

A posteriori to the calculation of the swimmer behavior, we determine the amplitudes of the anti-bending forces at each order $\epsilon^i$ by imposing \begin{equation}
\left(\vec{n}^\perp_I, -2 \vec{n}^\perp_I, \vec{n}^\perp_I, \vec{0}, \vec{0}, \vec{0} \right) \cdot \bm{\xi}^{\prime (i, \mathrm{all})} =^! 0,\ \  \left(\vec{0}, \vec{0}, \vec{0}, \vec{n}^\perp_{\II}, -2 \vec{n}^\perp_{\II}, \vec{n}^\perp_{\II} \right) \cdot \bm{\xi}^{\prime (i, \mathrm{all})} =^! 0,
\end{equation}
with $\bm{\xi}^{\prime (i, \mathrm{all})} = \sum_j q^j \bm{\xi}^{\prime (i, j)}$.  
Performing the steps described in this Appendix involves tedious algebraic operations, which is why we use the software Mathematica \cite{Mathematica2017} to perform the actual calculations.

\section{Details on the stroke-based perturbative calculation}
\subsection{Flow field of a single stroke-based linear three-sphere swimmer}
\label{ch:appendixC1}
To calculate the flow field produced by a single linear three-sphere swimmer, we firstly calculate the trajectory of each bead and from this the velocity field in the fluid. We hereby follow the calculations performed in \cite{GolestanianAjdari2008, AlexanderPooleyYeomans2009}. 

Similarly to the case of two interacting swimmers, the relation of the bead velocities and the forces necessary to enforce the prescribed swimming stroke is given by \eqref{eq:StokesLawCompact}. In the case of one swimmer aligned with the $y$-axis, a one-dimensional framework with $\bm{R} (t) = (y_1 (t), y_2 (t), y_3 (t))$ and $\bm{F} (t) = (F_1 (t), F_2 (t), F_3 (t))$ is sufficient to calculate the bead trajectories. In contrast to the case of two interacting swimmers, here the mobility matrix $\underline{\mu}(\bm{R})$ is determined already by the prescribed stroke, simplifying the calculation. 
With the two constraints resulting from the prescribed arm lengths, 
\begin{eqnarray}
    \dot{y}_2 (t) - \dot{y}_1 (t) = \dot{L}_1 (t), \nonumber \\
    \dot{y}_3 (t) - \dot{y}_2 (t) = \dot{L}_2 (t),
\end{eqnarray}
as well as the force-free condition, $F_1 (t) + F_2 (t) + F_3 (t) = 0$, the system of equations \eqref{eq:StokesLawCompact} is closed and can be solved using Mathematica \cite{Mathematica2017}. Here, the dot denotes the derivative with respect to time. 
With the prescribed stroke inserted, it is straight-forward to obtain the bead positions from the respective velocities by integrating with respect to time. 

The flow field produced by the swimmer, $\vec{u} (\vec{r}, t)$, is then given by \cite{AlexanderPooleyYeomans2009}
\begin{equation}
    \vec{u} (\vec{r}, t) = \sum_{i = 1}^3 \hat{T}(\vec{r} - \vec{x}_i (t)) \cdot \vec{F}_i (t), 
\end{equation}
with $\vec{x}_i (t) = (0, y_i (t))$ and $\vec{F}_i (t) = (0, F_i (t))$ the extensions of scalar quantities to two-dimensional vectorial quantities. 
Time-averaging and expanding the flow field with respect to $\vec{r}$ allows to decompose the flow field obtained into dipolar ($\sim 1/|\vec{r}|^2$), quadrupolar ($\sim 1/|\vec{r}|^3$), ... components. 

\subsection{Interaction of two stroke-based linear three-sphere swimmers}
\label{ch:appendixC2}
In this section, we elaborate on the details of the perturbative calculation for two interacting swimmer in the stroke-based model, which is outlined in section \ref{ch:strokeBasedModel}. A major complication in comparison to the calculation of the behavior of a single swimmer is that here the mobility matrix $\underline{\mu}(\bm{R})$ does not only depend on the arm lengths within each swimmer, which are prescribed, but also on the orientation of each swimmer and the relative positioning of both swimmers. Since the latter degrees of freedom are a priori unknown, we must use a perturbative approach in the swimmer separation $q$ to calculate their behavior. To speed up the calculation, we additionally perform an expansion also in the actuation amplitude $\xi$. 

With the constraints on the swimmers' linear shape as well as with the prescribed arm lengths, the positions of all beads are parametrized by the six remaining degrees of freedom by $\vec{R}_{I2}, \vec{R}_{\II 2}, \theta_I$ and $\theta_{\II}$. We expand each of those quantities in a power series with respect to $\xi$ and $q$: 
\begin{eqnarray}
\fl
\vec{R}_{I2} = \vec{R}_{I2}^\mathrm{init} + \sum_{i = 1}^{i_\mathrm{max}} \sum_{j = 0}^{j_\mathrm{max}} \xi^i q^j \vec{R}_{I2}^{(i, j)}, \ \vec{R}_{\II 2} = \vec{R}_{\II 2}^\mathrm{init} + \sum_{i = 1}^{i_\mathrm{max}} \sum_{j = 0}^{j_\mathrm{max}} \xi^i q^j \vec{R}_{\II 2}^{(i, j)}, \nonumber \\ \fl \theta_I = \theta_{I}^\mathrm{init} + \sum_{i = 1}^{i_\mathrm{max}} \sum_{j = 0}^{j_\mathrm{max}} \xi^i q^j \theta_I^{(i, j)}, \ \theta_{\II} = \theta_{\II}^\mathrm{init} + \sum_{i = 1}^{i_\mathrm{max}} \sum_{j = 0}^{j_\mathrm{max}} \xi^i q^j \theta_{\II}^{(i, j)},
\end{eqnarray}
where the first upper index in brackets corresponds to powers of $\xi$ and the second index to powers of $q$. Note that we omit the time-dependence of bead positions, swimmer orientations and swimmer arm lengths for the sake of brevity. $\vec{R}_{I2}^\mathrm{init}$, $\vec{R}_{\II 2}^\mathrm{init}$, $\theta_{I}^\mathrm{init}$ and $\theta_{\II}^\mathrm{init}$ denote the initial conditions for the two swimmers at $t = 0$. 
The positions of the remaining beads are then given for each swimmer $s$ by 
\begin{equation}
\vec{R}_{s1} = \vec{R}_{s2} - L_{s1} \vec{n}_s, \ \vec{R}_{s3} = \vec{R}_{s2} + L_{s2} \vec{n}_s, \ s \in \{I, \II \}, 
\end{equation}
with $\vec{n}_s = (\sin \theta_s, - \cos \theta_s)$. Using the Taylor expansion of the sine and cosine around the initial swimmer configuration, the positions of all beads can be expressed in terms of $\vec{R}_{I2}^{(i, j)}, \vec{R}_{\II 2}^{(i, j)}, \theta_I^{(i, j)}$ and $\theta_{\II}^{(i, j)}$. 

Applying a time-derivative to $\bm{R}$ allows to express the bead velocities $d/dt \, \bm{R}$ in dependence of $\dot{\vec{R}}_{I2}^{(i, j)}, \dot{\vec{R}}_{\II 2}^{(i, j)}, \dot{\theta}_I^{(i, j)}$ and $\dot{\theta}_{\II}^{(i, j)}$ as well as the swimmer orientations $\theta_I^{(i, j)}$ and $\theta_{\II}^{(i, j)}$.
Similarly, a Taylor expansion of the mobility matrix around the initial configuration of both swimmers allows to express $\underline{\mu}(\bm{R})$ in terms of $\vec{R}_{I2}^{(i, j)}, \vec{R}_{\II 2}^{(i, j)}, \theta_I^{(i, j)}$ and $\theta_{\II}^{(i, j)}$. 

With this expressions at hand, we are able to exploit \eqref{eq:StokesLawCompact} in order to solve for the forces $\bm{F}$. In order to keep the calculation simple, we expand \eqref{eq:StokesLawCompact} with respect to $q$ and find
\begin{align}
&\bm{F}^{(\mathrm{all}, 0)} = \left[ \underline{\mu}^{(\mathrm{all}, 0)} \right] \left( \frac{d}{dt} \bm{R} \right)^{(\mathrm{all}, 0)}, \nonumber \\
&\bm{F}^{(\mathrm{all}, 1)} = \left[ \underline{\mu}^{(\mathrm{all}, 0)} \right] \left[ \left( \frac{d}{dt} \bm{R} \right)^{(\mathrm{all}, 1)} - \underline{\mu}^{(\mathrm{all}, 1)} \bm{F}^{(\mathrm{all}, 0)} \right], \label{eq:solvedForForces}\\
&\bm{F}^{(\mathrm{all}, 2)} = \left[ \underline{\mu}^{(\mathrm{all}, 0)} \right] \left[ \left( \frac{d}{dt} \bm{R} \right)^{(\mathrm{all}, 2)} - \underline{\mu}^{(\mathrm{all}, 1)} \bm{F}^{(\mathrm{all}, 1)} - \underline{\mu}^{(\mathrm{all}, 2)} \bm{F}^{(\mathrm{all}, 0)}  \right], \nonumber \\
... \nonumber
 \end{align}
where the first index 'all' indicates that the corresponding quantity is not expanded with respect to $\xi$. A subsequent expansion of the right-hand sides of \eqref{eq:solvedForForces} allows to find expressions for the forces expanded with respect to both $q$ and $\xi$. 

The six remaining degrees of freedom, $\vec{R}_{I2}^\mathrm{init}$, $\vec{R}_{\II 2}^\mathrm{init}$, $\theta_{I}^\mathrm{init}$ and $\theta_{\II}^\mathrm{init}$, are then determined by the force-free and torque-free conditions \eqref{eq:forceFreeCond} and \eqref{eq:torqueFreeCond}. Expanding these equations by powers of $\xi$ and $q$, we notice that the force-free condition at each order $\xi^i q^j$ involves only $\bm{F}^{(i, j)}$ whereas the torque-free condition involves also components $\bm{F}^{(i', j')}$ with $i' \leq i$ and $j' \leq j$. 
Therefore, we employ again an ascending scheme through the orders in $\xi$ and $q$ given in this case by $q^0 \xi^1 \to q^0 \xi^2 \to q^1 \xi^1 \to q^1 \xi^2 \to q^2 \xi^1 \to q^2 \xi^2 \to q^3 \xi^1 \to ...$. Using this ascending scheme, all lower order components of the forces or $\vec{R}_{I2}, \vec{R}_{\II 2}, \theta_I, \theta_{\II}$ required at order $\xi^i q^j$ have already been computed. Consequently, the force-free and torque-free condition at each order $\xi^i q^j$, with all lower order components of $\bm{F}$ and $\vec{R}_{I2}, \vec{R}_{\II 2}, \theta_I, \theta_{\II}$ substituted by their explicit result, become algebraic equations for $\dot{\vec{R}}_{I2}^{(i, j)}, \dot{\vec{R}}_{\II 2}^{(i, j)}, \dot{\theta}_I^{(i, j)}$ and $\dot{\theta}_{\II}^{(i, j)}$. By solving these equations and time-integrating the results found we obtain the bead displacement at each order. 
A suitable decomposition into translational velocity along the swimmer axis, orthogonal to it and into rotational velocity yields the results for the different components of the interactions. Due to the increasing number of terms, we use Mathematica \cite{Mathematica2017} to perform the calculation. 

\section{Proportionality of $\xi^2$ average flow field and swimming velocity of a single swimmer}
\label{ch:appendixD}
We show that the amplitude of each $\xi^2$ component (dipolar, quadrupolar, ...) to the time-averaged flow field produced by the linear swimmer can be expressed as a purely geometric prefactor times $\int dt [\dot{\xi}_1(t) \xi_2(t) - \xi_1(t) \dot{\xi}_2(t)]$, which represents the area in the swimmer's configuration space enclosed by its trajectory.

As noted by Golestanian and Ajdari \cite{GolestanianAjdari2008}, the instantaneous velocity of each bead of a single swimmer along its axis, $v_i$, is linear in the velocities of the arm extensions and thus can be expressed as
\begin{equation}
v_i (t) = V^{i}_b \dot{\xi}_b (t) +  W^{i}_{bc} \dot{\xi}_b (t) \xi_c (t) + \mathcal{O}(\dot{\xi} \xi^2),
\label{eq:beadVelocity}
\end{equation}
with $b, c$ indices iterating over the swimmer arms and $V$, $W$ general geometrical prefactors. Repeated indices are summed over and the dot denotes derivation with respect to time. Assume the swimmer is oriented along the $y$-axis. 
Due to the overall linearity resulting from the Stokes equations, the fluid velocity $ \vec{u}(\vec{r}, t)$ at some arbitrary position $\vec{r}$ in the absolute coordinate system and time $t$ can then be written as a sum over all beads, 
\begin{equation}
    \vec{u}(\vec{r}, t) = \sum_i \hat{U}(\vec{r} - \vec{R}_i (t)) \cdot (0, v_i (t)),
    \label{eq:flowFieldGeneralFormula}
\end{equation}
with $(0, v_i(t))$ the 2D velocity vector of bead $i$ and $\hat{U}$ some tensor which depends on the vector connecting the position considered to the position of bead $i$.

From integration of \eqref{eq:beadVelocity} we know that up to order $\xi^1$, i.e. linear in either $\xi_b(t)$ or $\dot{\xi}_b (t)$, the position of each bead $i$ is a constant plus a term linear in $\xi_b (t)$ with $b = 1, 2$. Using a Taylor expansion around $\vec{r}$, thus also $\hat{U}(\vec{r} - \vec{R}_i)$ can be written, up to order $\xi^1$, as a constant plus a term linear in $\xi_b$. 
Hence, combining \eqref{eq:beadVelocity} and \eqref{eq:flowFieldGeneralFormula}, we find that all $\xi^2$ contributions to $\vec{u}(\vec{r})$ can then be written as 
\begin{equation}
\vec{u} (\vec{r}, t) = \vec{C}_{bc} (\vec{r}) \cdot \dot{\xi}_b \xi_c, 
\end{equation}
with $\vec{C}_{bc}(\vec{r})$ a prefactor which, besides the dependence on the position $\vec{r}$ in the absolute coordinate system, depends only on the geometry of the system. 
The only combination of $\dot{\xi}$ and $\xi$ which does not vanish when averaged over one stroke cycle is $\dot{\xi}_1 \xi_2 - \xi_1 \dot{\xi}_2$. This proofs our claim. 

We point out that in contrast to the velocity of each bead, the flow field at a fixed position in the absolute coordinate system does not depend linearly on $\dot{\xi}_b (t)$ when we consider terms of order higher than 2 in $\xi$. Since the swimmer self-propagates in the absolute coordinates and therefore relative to $\vec{r}$, the vector connecting $\vec{r}$ to $\vec{R}_i$ will at order $\xi^2$ also contain terms of the form $\int_0^t [\dot{\xi}_1 (t') \xi_2 (t') - \xi_1 (t') \dot{\xi}_2 (t')] dt'$. Therefore, the instantaneous flow field at order $\xi^3$ will also contain terms of the form $\int_0^t [\dot{\xi}_1 (t') \xi_2 (t') - \xi_1 (t') \dot{\xi}_2 (t')] dt' \dot{\xi}_b (t)$ and the expansion \eqref{eq:beadVelocity} is not valid for components of the instantaneous flow field.

\section*{References}

\bibliography{./Scheel201910.bib}

\bibliographystyle{iopart-num} 
\end{document}